\documentclass[
nofootinbib,
amsmath,amssymb,
aps,
prd
]{revtex4-1}
\usepackage{amsmath,amsfonts,amsthm,amssymb,mathtools}
\usepackage[dvips]{graphics,graphicx}
\usepackage[usenames,dvipsnames]{xcolor}
\definecolor{darkblue}{RGB}{0,0,196}
\definecolor{darkgreen}{RGB}{0,120,0}
\usepackage[colorlinks=true,linktocpage=true,linkcolor=darkblue,citecolor=red,urlcolor=darkblue]{hyperref}
\usepackage{cancel}
\usepackage{bbold}
\usepackage{multirow}
\usepackage{combelow}
\usepackage{longtable}
\usepackage{color}
\usepackage[normalem]{ulem}
\usepackage{bigints}
\usepackage{xparse}
\usepackage{physics}
\usepackage{verbatim}
\usepackage{minibox}
\usepackage{comment}
\usepackage{appendix}
\usepackage{marginnote}
\usepackage{graphicx}
\usepackage{dsfont}
\usepackage{slashed}
\usepackage[nice]{nicefrac}
\usepackage{tikz}
\usepackage{titlesec}

\def\eq{\mathrm{eq}}
\def\enp{h}
\newcommand{\n}{\nonumber \\ \linebreak}

\newcommand{\mcE}{{\mathcal{E}}}

\newcommand{\mcM}{{\mathcal{M}}}

\newcommand{\mcP}{{\mathcal{P}}}
\newcommand{\mcQ}{{\mathcal{Q}}}

\newcommand{\mcS}{{\mathcal{S}}}
\newcommand{\mcT}{{\mathcal{T}}}

\def\a{\alpha}
\def\b{\beta}

\def\d{\delta} 
\def\r{\rho}

\newcommand{\beq}{\begin{equation}}
\newcommand{\eeq}{\end{equation}}

\newcommand{\projkt}{\Phi}

\newcommand{\kt}{{k_\perp}}
\newcommand{\ktsup}[2][\perp]{k^{#2}_{#1}}
\newcommand{\ktsub}[2][\perp]{k_{#1#2}}
\def\eq{\mathrm{eq}}

\def\freq{\omega_{\rm lrf}}

\newcommand{\inv}[1]{\frac{1}{#1}}
\def\vs{v_s}
\def\vsw{v_\mathfrak{s}}

\def\fvort{\vartheta}
\def\kvort{\vartheta}
\def\elrf{equilibrium local rest frame}
\newcommand{\christoffel}[3]{\Gamma^{#1}_{#2#3}}

\newcommand{\revis}[1]{#1}
\begin{document}
\title{
    Semi-Classical Spin Hydrodynamics in Flat and Curved Spacetime: Covariance, Linear Waves, and Bjorken Background
    }
    \author{Annamaria Chiarini}
\email{chiarini@itp.uni-frankfurt.de}
    \author{Julia Sammet}
\email{sammet@itp.uni-frankfurt.de}
\author{Masoud Shokri}
\email{shokri@itp.uni-frankfurt.de}
\affiliation{Institut f\"ur Theoretische Physik, 
	Johann Wolfgang Goethe--Universit\"at,
	Max-von-Laue-Stra\ss e 1, D--60438 Frankfurt am Main, Germany}
\date{\today}

\begin{abstract}
We explore various aspects of semi-classical spin hydrodynamics, where hydrodynamic currents are derived from an expansion in the reduced Planck constant $\hbar$, incorporating both flat and curved spacetimes.
After establishing covariant definitions for angular momentum currents, we demonstrate that the conservation of the energy-momentum tensor requires modifications involving the Riemann curvature and the spin tensors.
We also revise pseudo-gauge transformations to ensure their applicability in curved spacetimes.

Key assumptions for semi-classical spin hydrodynamics are introduced, enabling studies without explicitly invoking quantum kinetic theory.
We derive and analyze the linearized semi-classical spin hydrodynamic equations, proving that spin and fluid modes decouple in the linear regime.
As a concrete example, we study the ideal-spin approximation in a dissipative fluid with shear viscosity. This analysis confirms our general result: the damping of spin waves is governed solely by spin relaxation time coefficients, independent of linear fluid perturbations.

We also examine the Gibbs stability criterion and reveal its limitations at first order in $\hbar$, signaling the inherent anisotropy of the equilibrium state, which remains unaddressed in current semi-classical spin hydrodynamics formulations.
Finally, within a conformal Bjorken flow background and using the slow-roll approximation attractor for the fluid sector, we show that the relaxation of the spin potential is governed by spin relaxation time coefficients, mirroring the damping behavior of spin waves in the linear regime.
\end{abstract}

\maketitle

\section{Introduction and summary}\label{sec:intro}

Hydrodynamics is an effective theory describing the low-frequency, long-wavelength dynamics of many-body systems based on the conservation of charges \cite{rezzolla2013relativistic, rischke_denicol_book, Florkowski:2017olj}.
A fundamental assumption of relativistic hydrodynamics is the concept of local thermodynamic equilibrium (LTE), a local map between a fluid cell---defined as a macroscopically infinitesimal domain---and a fictitious \textit{global} thermodynamic equilibrium state, where the system’s entropy is maximized \cite{Israel:1979wp}.
For such a map to hold, the characteristic size of a fluid cell, denoted as  $L_{\rm hydro}$, must satisfy the hierarchy  $l_{\rm micro} \ll L_{\rm hydro} \ll L_{\rm sys}$, where  $l_{\rm micro}$ is the microscopic length scale, and $L_{\rm sys}$ represents the typical size of the system. 
The ratio between the microscopic and hydrodynamic scales, $\mathrm{Kn} \sim l_{\rm micro}/L_{\rm hydro}$, is known as the Knudsen number. 
A small Knudsen number, $\mathrm{Kn} \ll 1$, is required for the validity of the hydrodynamic description.

In the absence of an inherent spatial anisotropy, the energy-momentum tensor in global equilibrium assumes the \textit{perfect fluid} form:
\begin{equation}\label{eq:perfect-fluid}
    T^{\mu\nu} = \varepsilon u^\mu u^\nu - P \Delta^{\mu\nu}\;,
\end{equation}
where $\varepsilon$ is the energy density, $u_\mu$ is the fluid four-velocity, $P$ is the pressure, and $\Delta^{\mu\nu} = g^{\mu\nu}-u^\mu u^\nu$,  in the mostly-minus sign convention used in this paper, projects every vector onto the plane orthogonal to $u^\mu$. 
Importantly, while Eq.\ \eqref{eq:perfect-fluid} resembles the form of a perfect fluid’s energy-momentum tensor, in global equilibrium, the reduction to this form does not stem from the vanishing of the Knudsen number but rather from entropy being maximized.
For uncharged dissipative fluids, the energy-momentum tensor is derived by expanding around the reference global equilibrium state.
The resulting dissipative corrections to Eq.\ \eqref{eq:perfect-fluid} were obtained employing various approaches, including the Israel-Stewart (IS) theory \cite{ISRAEL1976310,stewart1977transient,Israel:1979wp}, its systematic extension based on kinetic theory, known as the Denicol-Niemi-Molnar-Rischke (DNMR) theory \cite{Denicol:2012cn,Denicol:2012es}, and the most recent Bemfica-Disconzi-Noronha-Kovtun (BDNK) theory \cite{Bemfica:2020zjp,Kovtun:2012rj}.
Regardless of the approach, we refer to the expansions used by them as the \textit{classical hydrodynamic expansion}.

Recently, the observation of polarization of $\Lambda$ hyperons in heavy-ion collisions \cite{STAR:2017ckg, STAR:2018gyt, ALICE:2019aid} has sparked a growing interest in \textit{spin hydrodynamics} \cite{Florkowski:2017ruc, Shi:2020htn, Speranza:2020ilk, Bhadury:2020cop, Singh:2020rht, Bhadury:2021oat, Peng:2021ago, Hu:2021pwh, Montenegro:2018bcf,Florkowski:2019qdp, Montenegro:2020paq, Gallegos:2021bzp, Hattori:2019lfp, Fukushima:2020ucl, Singh:2021man,Li:2020eon, She:2021lhe, Hongo:2021ona,Singh:2022ltu,Ambrus:2022yzz, Daher:2022xon, Weickgenannt:2022zxs, Weickgenannt:2022jes, Bhadury:2022ulr, Torrieri:2022ogj, Wagner:2022gza, Weickgenannt:2022qvh, Daher:2022wzf, Xie:2023gbo, Shi:2023sxh, Biswas:2023qsw, Becattini:2023ouz, Weickgenannt:2023btk, Weickgenannt:2023bss, Kiamari:2023fbe, Drogosz:2024gzv, Weickgenannt:2024esg, Wagner:2024fhf, Wagner:2024fry,Singh:2024cub,Wang:2021ngp,Wang:2021wqq,Wang:2024afv}. 
Spin hydrodynamics extends standard relativistic hydrodynamics by treating the total angular momentum current $J^{\lambda\mu\nu}$ as an independent conserved charge density current, where $J^{\lambda\mu\nu}$ is a rank-3 tensor antisymmetric in its last two indices.
This rank-3 tensor is typically decomposed into orbital and spin parts:
\begin{equation}\label{eq:angular-momentum-rank3}
    J^{\lambda\mu\nu} = L^{\lambda\mu\nu} + \mcS^{\lambda\mu\nu}\;,
\end{equation}
where $L^{\lambda\mu\nu}$ is the orbital angular momentum and $\mcS^{\lambda\mu\nu}$ is the spin tensor.
Substituting the definition $L^{\lambda\mu\nu} =  2 T^{\lambda[\nu}x^{\mu]}$, in Cartesian coordinates, and the conservation of the energy-momentum tensor $\partial_\mu T^{\mu\nu}=0$ in the conservation of total angular momentum tensor, $\partial_\lambda J^{\lambda\mu\nu} = 0$, leads to the spin tensor equation of motion:
\begin{equation}\label{eq:spin-dynamics-cartesian}
    T^{[\mu\nu]} =  -\frac{1}{2}\partial_\lambda \mcS^{\lambda\mu\nu} \;.
\end{equation}
Here, the antisymmetrization of a rank-2 tensor $T$ is defined as $T^{[\mu\nu]}\equiv \tfrac{1}{2}(T^{\mu\nu}-T^{\nu\mu})$.

Taking the angular momentum as an independent charge naturally highlights the relevance of its conjugate quantity, angular velocity \cite{Landau_vol5}, which is crucial because global equilibrium states can exhibit rigid rotation, inducing equilibrium gradients \cite{Shokri:2023rpp}.
This issue was briefly noted in Ref.\ \cite{Israel:1979wp} and is elaborated in this work.
To prevent the development of large equilibrium gradients, a rigidly rotating equilibrium state must satisfy the condition of slow rotation.
Violating this condition results in arbitrarily large accelerations in some regions of spacetime, inducing anisotropy that necessitates a revision of the hydrodynamic currents in equilibrium.

The definition of orbital angular momentum, $L^{\lambda\mu\nu} =  2 T^{\lambda[\nu}x^{\mu]}$, is specific to Cartesian coordinates and is not covariant under general coordinate transformations.
We address this issue, by redefining the orbital and total angular momentum in a fully covariant manner:
\begin{equation}
     J^{\lambda r} = L^{\lambda r} +  \mcS^{\lambda r} \;,
\end{equation}
where the orbital and spin contributions are given by 
\begin{equation}
    L^{\lambda r} \equiv - T^{\lambda\nu}K^r_\nu\;, \qquad  \mcS^{\lambda r} \equiv \frac{1}{2} \mcS^{\lambda\mu\nu}D_{[\nu}K^r_{\mu]}\;.
\end{equation}
Here, $K^r_\mu$ represents the Killing vector field that generates rotations, and $D$ is the covariant derivative.
This covariant formulation requires the equations of motion to be modified as
\begin{equation}
    D_\mu T^{\mu\nu}=-\frac{1}{2}R^{\nu}{}_{\a\b\gamma}\mcS^{\a\b\gamma}\;,\qquad T^{[\mu\nu]}=-\frac{1}{2}D_\lambda \mcS^{\lambda\mu\nu}\;,
\end{equation}
where $R^\sigma_{\r\mu\nu} \equiv 2\left(\partial_{[\mu}\christoffel{\sigma}{\nu]}{\rho}
+\christoffel{\sigma}{[\mu}{\beta}\christoffel{\beta}{\nu]}{\r}\right)$ is the Riemann tensor.

\revis{The curvature-induced modification of the equations of motion for polarized media, assuming test particles obey the Mathisson–Papapetrou–Dixon equations \cite{1951RSPSA.209..248P,1970RSPSA.314..499D,Mathisson:1937zz}, was derived by Israel in Ref.\ \cite{israel1973electrodynamics}.
Hehl \ obtained the modified equation of motion \cite{HEHL197655,RevModPhys.48.393} for Dirac fermions, using the variational principle within the Einstein–Cartan framework.
More recently, it has been applied to spin hydrodynamics \cite{Gallegos:2021bzp,Gallegos:2022jow}, where torsion serves as a power-counting tool.
In contrast, our analysis is formulated in a torsionless background and guided by the principle that the equations of motion must be \textit{pseudo-gauge covariant}, while conserved quantities remain \textit{pseudo-gauge invariant}.
Accordingly, our definition of total angular momentum is invariant under pseudo-gauge transformations, provided the definition of the transformation itself is appropriately generalized to account for curvature-dependent contributions.
These subtleties disappear in flat spacetime, where the generalized transformations reduce to their standard form.}

In the classical hydrodynamic expansion, the fluid's quantum nature is reflected solely in the transport coefficients, while the constitutive relations remain agnostic to the underlying microscopic theory. 
Spin hydrodynamics, however, incorporates spin---of an inherently quantum nature---into the macroscopic theory, requiring an additional expansion in quantum corrections.
To address this, we naturally employ the semi-classical expansion scheme, where all quantities are systematically expanded in $\hbar$, with the equations, rather than the quantities, truncated at the first order.
The resulting formulation, termed \textit{semi-classical spin hydrodynamics}, is inspired by quantum kinetic theory results \cite{Weickgenannt:2019dks,Weickgenannt:2022jes,Wagner:2024fry}, without explicitly relying on them.

At first order in $\hbar$, there is no back-reaction from the spin sector to the fluid dynamics, allowing solutions to standard hydrodynamics equations of motion to serve as inputs for the spin tensor equations of motion.
Leveraging this fact, spin relaxation was studied using the simplest solution to the fluid equations in Ref.\ \cite{Wagner:2024fhf}: the hydrostatic equilibrium.
In this work, we extend the analysis to two other cases: \textbf{linearized hydrodynamics} and \textbf{conformal Bjorken flow}.
We demonstrate that spin and fluid waves decouple at this order, and the results from linearized hydrodynamics match those of the hydrostatic case, showing that linear fluid perturbations do not alter the conclusions of Ref.\ \cite{Wagner:2024fhf}.

Furthermore, we study the thermodynamic stability by assessing the Gibbs stability criterion \cite{Hiscock:1983zz,Gavassino:2021cli}, which relies on maximizing entropy subject to conserved charges.
However, when applied to semi-classical spin hydrodynamics, it encounters limitations.
Specifically, we demonstrate that truncating the hydrodynamic equations at first order in $\hbar$ results in stability conditions that pertain exclusively to the fluid sector.

The Bjorken flow \cite{Hwa:1974gn,Bjorken:1982qr} has long been employed as a simple toy model for heavy-ion collisions in various contexts, including spin hydrodynamics theories derived from classical kinetic theory and entropy current analysis \cite{Florkowski:2019qdp,Wang:2021ngp,Wang:2024afv}.
Here, we study semi-classical spin hydrodynamics within the conformal Bjorken flow background. 
For this purpose, we revisit the consequences of conformal invariance in spin hydrodynamics, previously explored in Ref.\ \cite{Singh:2020rht}. 
{Contrary to the conclusions of Ref.\ \cite{Singh:2020rht}, which argued that conformal invariance requires a traceless energy-momentum tensor and a totally antisymmetric spin tensor---effectively eliminating pseudo-gauge freedom---we show that such constraints are not necessary.
We postulate Weyl covariance of the conserved charge currents and derive the resulting transformation laws. 
Our findings are consistent with the variational approach of Ref.\ \cite{Gallegos:2022jow}.
Assuming that conformal symmetry holds only in the classical limit $\hbar \to 0$, we employ the attractor solution for the fluid sector to derive the evolution of the spin potential.
We demonstrate that, even in this nonlinear regime, its relaxation timescale closely tracks the damping of spin waves in a hydrostatic background.}

This paper is organized as follows.
In Sec.\ \ref{sec:spin-hydro}, we discuss the covariance of spin hydrodynamics. Section \ref{sec:semi-classical} introduces the key assumptions of semi-classical hydrodynamics and reviews the ideal-spin approximation.
In Sec.\ \ref{sec:linear-spin-hydro}, we present a general result for linear spin hydrodynamics, followed by an examination of the Gibbs stability criterion in the context of semi-classical spin hydrodynamics in Sec.\ \ref{sec:entropy}.
Section \ref{sec:example} illustrates our findings from Sec.\ \ref{sec:linear-spin-hydro} with a concrete example.
In Sec.\ \ref{sec:bjorken}, we explore the implications of conformal Bjorken flow for spin hydrodynamics.
Finally, the paper concludes in Sec.\ \ref{sec:conclusion}.
Details of lengthy calculations and additional results are provided in the appendices.

\paragraph{Notations and conventions} 
For the reader's reference, we define notations and conventions used throughout this work.
Throughout this work, we use natural units, setting $c = k_B = 1$, while retaining the reduced Planck constant $\hbar$ to track the order of quantum effects. 
The metric sign convention is mostly minus, $\eta_{\mu\nu} = \mathrm{diag}(1, -1, -1, -1)$, and the convention for the Levi-Civita tensor in Cartesian coordinates is $\epsilon_{0123} = -\epsilon^{0123} = -1$.
The scalar product of two four-vectors $a^\mu$ and $b^\mu$ is denoted by $a \cdot b \equiv a^\mu b_\mu$. 
Antisymmetrization and symmetrization of a rank-2 tensor $A$ are defined as $A^{[\mu\nu]} \equiv \frac{1}{2}(A^{\mu\nu} - A^{\nu\mu})$ and $A^{(\mu\nu)} \equiv \frac{1}{2}(A^{\mu\nu} + A^{\nu\mu})$, respectively.
The comoving derivative of a tensor $X$ is denoted by $\dot{X} \equiv u \cdot \partial X$, where $u^\mu$ is the four-velocity normalized as $u \cdot u = 1$. The projection of a vector $V^\mu$ is written as $V^{\langle\mu\rangle} \equiv \Delta^{\mu\nu}V_\nu$. 
For a scalar quantity $X$, the projected gradient is defined via $\nabla_\mu X \equiv \Delta_{\mu}^{\rho}D_\rho X$, where $D_\mu$ denotes the covariant derivative. For simplicity, we also adopt the semicolon notation $A_{\a;\b} \equiv D_\b A_\a$.
The traceless symmetric rank-4 projector is $\Delta^{\mu\nu}_{\alpha\beta} \equiv \Delta^{(\mu}_\alpha \Delta^{\nu)}_\beta - \frac{1}{3} \Delta^{\mu\nu} \Delta_{\alpha\beta}$, and a projected rank-2 tensor is denoted as $A^{\langle\mu\nu\rangle} \equiv \Delta^{\mu\nu}_{\alpha\beta} A^{\alpha\beta}$.
The four-gradient of $u_\mu$ decomposes as
\begin{equation*}
D_\mu u_\nu = u_\mu a_\nu + \frac{1}{3} \theta \Delta_{\mu\nu} + \sigma_{\mu\nu} + \omega_{\mu\nu},
\end{equation*}
where $a_\nu \equiv u^\a  D_\a u_\nu$ is the four-acceleration, $\theta = D \cdot u$ is the expansion scalar, $\sigma_{\mu\nu} = \Delta^{\a\b}_{\mu\nu}D_\a u_\b$ is the shear tensor, and $\omega_{\mu\nu} \equiv \nabla_{[\mu} u_{\nu]}$ is the fluid vorticity tensor.
The Hodge dual of a rank-2 tensor $A^{\mu\nu}$ is defined as $\tilde{A}^{\mu\nu} \equiv \frac{1}{2} \epsilon^{\mu\nu\a\b}A_{\a\b}$.

\section{Foundations of spin hydrodynamics}\label{sec:spin-hydro}
In this section, we first develop a covariant formulation of spin hydrodynamics in flat spacetime.
In particular, we discuss how the generators of rotation are used to define the total angular momentum covariantly.
Next, we discuss how the intensive parameters of the \textit{environment} and the spacetime geometry determine the so-called thermal Killing vector and, thus, the state of global equilibrium.
We also briefly review the concept of local thermodynamic equilibrium as a mapping between the fluid and space of possible equilibrium states.
Finally, we demonstrate that these definitions can be extended to curved spacetime, provided the equations of motion are reformulated to ensure that the total angular momentum remains pseudo-gauge independent and conserved.
This reformulation agrees with the results of Ref. \cite{Gallegos:2022jow}.

\subsection{Covariant conserved charge currents in flat spacetime}\label{sec:charge-currents}

The equations of hydrodynamics are rooted in the conservation of charges, expressed as surface integrals of currents $J^{\mu I}$ over arbitrary Cauchy hypersurfaces $\Sigma$:
\begin{equation}\label{eq:cons_charges}
      Q^I = \int_{\Sigma} \dd{\Sigma_\mu} J^{\mu I}\;,
\end{equation}
where $I$ is the charge index, and $D_\mu J^{\mu I} = 0$.
Here, $\dd{\Sigma_\mu} \equiv n_\mu \dd{\Sigma}$, with $n_\mu$ being the timelike unit vector normal to $\Sigma$, and $\dd{\Sigma}$ being the surface element \cite{Poisson:2009pwt}.
In the absence of internal symmetries, i.e., for a so-called \textit{uncharged fluid}, conserved charges arise solely from spacetime symmetries, generated by the set of all independent Killing vectors $\{K^{I}\}$.
For a symmetric energy-momentum tensor $T^{\mu\nu}$ the divergence of $T^{\mu\nu}K^I_\nu$ vanishes, making it a conserved charge current.
This follows from the Killing condition $K^{I}_{(\a;\b)} =0$ and the conservation of energy-momentum tensor:
\begin{equation}\label{eq:em-cons}
    D_\mu T^{\mu\nu} = 0\;.
\end{equation}
However, when $T^{\mu\nu}$ is not symmetric, but remains conserved, this definition of conserved charge currents must be revised.
In this case, the energy-momentum tensor $T^{\mu\nu}$ can be decomposed into symmetric and antisymmetric parts as
\begin{equation}\label{eq:emt-sym-asym}
    T^{\mu\nu} = T^{(\mu\nu)} + T^{[\mu\nu]}\;,
\end{equation}
where the antisymmetric part satisfies 
\begin{equation}\label{eq:spin-dynamics-postulate}
   T^{[\mu\nu]} =  -\frac{1}{2}D_\lambda \mcS^{\lambda\mu\nu} \;.
\end{equation}
This relation, referred to as the \textit{spin dynamics} equation, generalizes its Cartesian counterpart \eqref{eq:spin-dynamics-cartesian} by replacing the partial derivative with a covariant derivative.
To account for the antisymmetric part of $T^{\mu\nu}$, we define the following set of currents:
\begin{equation}\label{eq:cons_current}
    J^{\mu I} \equiv s_K \left(T^{\mu\nu}K_\nu^I - \frac{1}{2}\mcS^{\mu\a\b}K^I_{[\a;\b]}\right)\;,
\end{equation}
where $s_K=1(-1)$ for timelike (spacelike) Killing vectors.
Using the Killing condition, the energy-momentum conservation \eqref{eq:em-cons}, the spin dynamics equation \eqref{eq:spin-dynamics-postulate}, and noting that $K^I_{\a;\b\gamma} = 0$ in flat spacetime, we obtain 
\begin{equation}\label{eq:cons_current-div}
    D_\mu J^{\mu I} = 0\;.
\end{equation}
This confirms that the currents $J^{\mu I}$ represent the fluxes of conserved charges $Q^I$, as defined in Eq.\ \eqref{eq:cons_charges}.

The energy-momentum and spin tensors are known to be non-unique and can be redefined using the so-called pseudo-gauge transformations \cite{HEHL197655,Speranza:2020ilk,Buzzegoli:2024mra},
\begin{equation}\label{eq:psgt}
    T^{\mu\nu}{}' = T^{\mu\nu} + D_\lambda Z^{\lambda\mu\nu}\;,\qquad \mcS^{\lambda\mu\nu}{}' = \mcS^{\lambda\mu\nu} - \Phi^{\lambda\mu\nu} - D_\rho \Xi^{\mu\nu\lambda\rho}\;,
\end{equation}
where the \textit{superpotential} $\Phi^{\lambda\mu\nu}$ is an arbitrary tensor antisymmetric in the last two indices, and $\Xi^{\mu\nu\lambda\rho}$ is an arbitrary tensor antisymmetric in the first and second pairs of indices.
The tensor $ Z^{\lambda\mu\nu}$ is defined as
\begin{equation}
    Z^{\lambda\mu\nu} \equiv \inv{2}\left( \Phi^{\lambda\mu\nu} - \Phi^{\mu\lambda\nu} - \Phi^{\nu\lambda\mu}\right)\;. 
\end{equation}
Inserting pseudo-gauge transformation \eqref{eq:psgt} into Eq.\ \eqref{eq:cons_current}, the transformed $J^{\mu I}{}'$ is found to be:
\begin{equation}
    J^{\mu I}{}' =J^{\mu I} - s_K D_\a A^{[\mu\a]}\;,
\end{equation}
where 
\begin{equation}
    A^{[\mu\a]} = Z^{\mu\a\nu}K^I_\nu - \frac{1}{2}D_\sigma \left( \Xi^{\b\sigma\mu\a}K_\b \right)\;.
\end{equation}

Since $A^{[\mu\a]}$ is a rank-2 antisymmetric tensor, the integral $\int \dd{\Sigma}_\mu D_\a A^{[\mu\a]}$ can be expressed as a boundary term by Stokes' theorem.
Therefore, with proper boundary conditions, the charge $Q^I$, as defined in Eq.\ \eqref{eq:cons_charges}, remains invariant under pseudo-gauge transformations.

\subsection{Covariant angular momentum in flat spacetime}\label{sec:angular-momentum}
Let us now insert the three generators of rotations $K^r$, with $r=1,2,3$, as the Killing vectors appearing in Eq.\ \eqref{eq:cons_current}.
These Killing vectors generate three currents, which represent the fluxes of the total angular momentum components:
\begin{equation}\label{eq:tot-angular-momentum}
    J^{\lambda r} = L^{\lambda r} +  \mcS^{\lambda r} \;,
\end{equation}
where, setting $s_K = -1$ in  Eq.\ \eqref{eq:cons_current} as $K^r$ are spacelike, the covariant forms of the orbital and spin angular momentum currents in flat spacetime, respectively, are
\begin{equation}\label{eq:spin-orbit-cov}
     L^{\lambda r} \equiv - T^{\lambda\nu}K^r_\nu\;, \qquad  \mcS^{\lambda r} \equiv \frac{1}{2} \mcS^{\lambda\mu\nu}D_{[\nu}K^r_{\mu]}\;.
\end{equation}
Substituting the explicit expression for the rotational Killing vector in Cartesian coordinates,
\begin{equation}\label{eq:gen-rot}
K^r = \epsilon^{rij}x^i\pdv{x^j}\;,
\end{equation}
Eq.\ \eqref{eq:tot-angular-momentum} can be recast in terms of rank-3 tensors of Eq.\ \eqref{eq:angular-momentum-rank3}, as illustrated in Appendix \ref{app:slow-rotation}.
The components of the total angular momentum are obtained by inserting $J^{\lambda r}$ in Eq.\ \eqref{eq:cons_charges},  yielding a set of three scalars
\begin{equation}\label{eq:total-angular-momentum-integrated}
    J^{r} = \int_{\Sigma}\dd{\Sigma}_\lambda J^{\lambda r}\;. 
\end{equation}
This is the covariant analog of the often-used relation in Cartesian coordinates
\begin{equation}
    J^{\mu\nu} = \int_{\Sigma}\dd{\Sigma}_\lambda J^{\lambda\mu\nu}\;.
\end{equation}
\subsection{Global and local thermodynamic equilibrium}\label{sec:geq-leq}
We now review the concept of \textit{global thermodynamic equilibrium (GTE)}, a state where the total entropy is maximized subject to the constraint of known conserved charges \cite{Balian2007}.
Here, the total entropy refers to the entropy of the body and its environment, with which the body is in equilibrium. 
Together, the body and its environment form a closed system. 
For a fluid, the body might refer to the fluid itself and the environment as its surrounding medium.
Alternatively, the body may be a portion of the fluid and the remaining parts of the fluid serve as the environment.

Describing global equilibrium states in relativistic systems requires identifying spacetime symmetries that generate conserved charges, which, as noted earlier, are encoded in the Killing vectors $K^I$.
These Killing vectors can be combined to define the ``thermal Killing vector'' \cite{Gavassino:2023qnw},
\begin{equation}\label{eq:thermal_killing}
    \beta^\star = \lambda^\star_I K^I\;,
\end{equation}
where the superscript $\star$ denotes global equilibrium, and the Einstein summation is assumed for the index $I$. The coefficients $\lambda^\star_I$ are the equilibrium intensive parameters:
\begin{equation}\label{eq:lambda-coeffs}
    \lambda^\star_I = \pdv{S}{Q^I}\Big|_{\rm GTE}\;,
\end{equation}
with $S$ representing the fluid entropy.
In equilibrium, these intensive parameters are identical for the body and the environment.
The thermal Killing vector $\beta^\star$ is timelike, i.e.,
\begin{equation}\label{eq:thermal_causality}
    \beta^\star_\mu \beta^{\star\mu} > 0\;.
\end{equation}
Consequently, the thermal Killing vector determines the fluid velocity $u_\mu$ and temperature $T$ uniquely as
\begin{equation}
    u_\mu = \frac{\beta^\star_\mu}{\sqrt{\beta^\star\cdot \beta^\star}}\;,\qquad  T = \inv{\sqrt{\beta^\star\cdot \beta^\star}}\;.
\end{equation}
Other thermodynamic quantities follow from thermodynamic identities and the fluid's equation of state.
In global equilibrium, $T^{\mu\nu}$ automatically satisfies the energy-momentum conservation \eqref{eq:em-cons}, rendering the equation trivial.

Since $\beta^\star$ is a Killing vector, the symmetric part of its gradient, $D_{(\mu}\beta^{\star}_{\nu)}$, vanishes.
The antisymmetric part, however, defines the so-called thermal vorticity in equilibrium
\begin{equation}\label{eq:tvort}
	\varpi^\star_{\mu\nu} \equiv -D_{[\mu}\b^\star_{\nu]}\;.
\end{equation}
As discussed in Ref.\ \cite{Shokri:2023rpp}, thermal vorticity serves as the differentiator between homogeneous ($\varpi^\star=0$) and inhomogeneous ($\varpi^\star\neq0$) equilibrium configurations.
The thermal vorticity is a rank-2 antisymmetric tensor, analogous to the Faraday tensor, and can be decomposed into magnetic and electric parts.
In global equilibrium, this decomposition takes the form
\begin{equation} \label{eq:vorticity}
\varpi^\star_{\mu \nu} = \frac{2}{T} a_{[\mu} u_{\nu]}
+ \frac{1}{T} \epsilon_{\mu \nu \alpha \beta}
\kvort^\alpha u^\beta\;,
\end{equation}
where the electric component $a_\mu \equiv u^\a D_\a u_\mu =  T \varpi_{\mu\nu}u^\nu$ represents the acceleration and the magnetic component $\kvort^\mu \equiv -\tfrac{1}{2} \epsilon^{\mu\nu\a\b}u_\nu \nabla_\a u_\b \equiv \tfrac{1}{2}T \epsilon^{\mu\nu\a\b} u_\nu \varpi_{\a\b}$ is the fluid vorticity vector.
As shown in Appendix \ref{app:slow-rotation}, the rotational components of thermal vorticity are directly related to intensive parameters corresponding to the angular momentum.

If the total energy $E$ is the only nonvanishing conserved charge in a fluid in equilibrium with a heat bath, the fluid must be at rest relative to the heat bath. 
In this case, the only intensive parameter is $\lambda^\star_0 = 1/T$, and the thermal Killing vector reduces to
\begin{equation}\label{eq:bstar-rest}
\beta^\star = b^\star \equiv \inv{T}\pdv{t}\;.
\end{equation}
For other equilibrium configurations, the thermal Killing vector takes the form
\begin{equation}\label{eq:cov_beta_star}
    \beta^\star = b^\star - \lambda^\star_i K^i\;,
\end{equation}
where we have defined $I=(0,i)$, with the index $i$ labeling the spacelike Killing vectors.
In flat spacetime, $i=1,\cdots,9$, corresponding to the generators of the Poincar\'{e} algebra excluding time translations.
However, three of these Killing vectors, representing spatial translations, can be eliminated through global Lorentz boosts, leaving six independent Killing vectors.
This number matches the independent degrees of freedom in the thermal vorticity.
As a result, Eq.\ \eqref{eq:cov_beta_star} represents the covariant form of the standard relation in Cartesian coordinates \cite{DeGroot:1980dk},
\begin{equation}\label{eq:sr_beta_star}
    \beta^\star_\mu = b^\star_\mu + \varpi^\star_{\mu\nu} x^\nu\;.
\end{equation}
The coefficients $\lambda_r$, associated with the generators of rotations appearing in Eq.\ \eqref{eq:cov_beta_star}, are of the form $\Omega_r / T_0$, where $\Omega_r$ are constants with dimension of energy, and $T_0$ is the temperature at the center of rotation. 
The causality constraint introduces a characteristic length scale $R\sim 1/\Omega_r$.
For further details, see Appendix \ref{app:slow-rotation}.
\begin{figure}
    \centering
    \begin{tikzpicture}

    \draw[thick] (-2, 0) rectangle (2, 2) node[anchor=west,pos=0.75] {\large $\mathcal{M}$};

    \node[circle, fill=black, inner sep=1pt, label=above:$x$] (x) at (0, 1.5) {};

    \draw[thick] (-2, -1) -- (2, -1) node[above right] {\large $\Sigma_{\rm EQ}$};

      \node[circle, fill=black, inner sep=1pt, label=below:{$e=(\varepsilon, u)$}] (mapped) at (1, -1) {};

    \draw[dashed, ->] (x) -- (mapped) node[midway, right] {};

\end{tikzpicture}
    \caption{Local thermodynamic equilibrium as a map between the fluid $\mcM$ and the set of all possible homogeneous equilibrium states $\Sigma_{\rm EQ}\,.$}
    \label{fig:lte-hydro}
\end{figure}
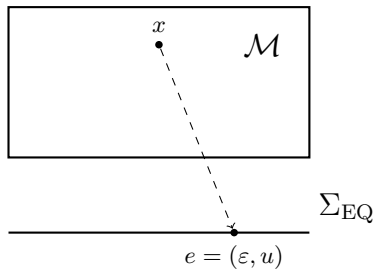

Before proceeding further, we briefly comment on the concept of LTE.
Following Ref.\ \cite{Israel:1979wp}, we define LTE as the mapping of each fluid cell at a point $x$ to a point $e$ in the set of all possible (global) equilibrium states of the same fluid, denoted as $\Sigma_{\rm EQ}$.
For an uncharged fluid, $\Sigma_{\rm EQ}$ forms a ten-dimensional space: the first four dimensions correspond to four-velocity $u$ and energy density $\varepsilon$, while the remaining six represent the components of thermal vorticity.
If angular momentum is neglected and only homogeneous equilibrium states are considered, the relevant subset of $\Sigma_{\rm EQ}$ reduces to a four-dimensional space, represented by the horizontal line in Fig.\ \ref{fig:lte-hydro}.
In this case, the LTE mapping involves two steps: (1) defining a local rest frame at each point, which determines $u(x)$, and (2) identifying the corresponding equilibrium state $e$, characterized by $u(e)=u(x)$ and the same energy density $\varepsilon(e)$ as in the fluid's local rest frame.
The first step is commonly referred to as a choice of \textit{a hydrodynamic frame} \cite{Kovtun:2012rj}, while the second step entails the \textit{matching condition}
\begin{equation}
    \varepsilon(e) = T^{\mu\nu}(x)u_\mu(x) u_\nu(x)\;.
\end{equation}
Thus, the LTE is a map between the fluid and the space of equilibrium states.
The term ``local'' highlights the point-by-point nature of this mapping, while each fictitious mapped equilibrium state corresponds to a \textit{global} equilibrium state. 

This map is uniquely defined only for a perfect fluid. 
Thus, a perfect fluid can be described as \textit{being in} local thermodynamic equilibrium, as it can be unambiguously mapped to an equilibrium state at every point.
In dissipative fluids, by contrast, the mapping is ambiguous, depending on the choice of a hydrodynamic frame.\footnote{The BDNK theory does not rely on matching conditions; thus, the LTE concept requires considerations beyond the scope of this work.}
The currents are then obtained by expanding around the mapped---or reference---global equilibrium state.
At equilibrium, the equations of motion are satisfied by construction, meaning that the zeroth-order terms in this expansion must automatically satisfy equations of motion under equilibrium conditions, and the higher-order contributions must vanish.

It is important to note that, in global equilibrium, the absence of higher-order contributions to the hydrodynamic currents is independent of the microscopic nature of the fluid.
Instead, as explained earlier, the state is dictated solely by the geometry of spacetime and the intensive parameters of the environment.
For a perfect fluid, however, the hydrodynamic currents contain only zeroth-order terms because the transport coefficients associated with dissipation vanish.

In spin hydrodynamics, where the angular momentum is treated as an independent charge, the concept of LTE requires additional considerations. 
In this work, we adopt the semi-classical expansion, with the LTE mapping discussed here serving as a valid approximation in the classical limit, i.e., at zeroth order in $\hbar$.
To keep the scope of this manuscript focused, we defer these further considerations to future work.

\subsection{Generalization to curved spacetime}

It is natural to ask whether the covariant framework presented here extends to curved spacetimes.
We limit our treatment to spacetimes, or submanifolds thereof, where a timelike Killing vector exists, allowing for the definition of a thermal Killing vector. 
Additionally, the spacetime must possess at least one Killing vector associated with rotational symmetry.

Under these conditions, the covariant definition of angular momentum can be generalized to torsionless curved spacetimes. 
To do so, we postulate that the divergence of the currents $J^{\mu I}$, defined as in Eq.\ \eqref{eq:cons_current}, vanishes.
To compute this divergence, we use the Killing equation, decompose the energy-momentum tensor into the symmetric and antisymmetric parts, Eq.\ \eqref{eq:emt-sym-asym}, and utilize the following identity (see, e.g., Refs.\ \cite{Poisson:2009pwt,Shokri:2023rpp}):
\begin{equation}
    K^I_{\a;\b\gamma} =  R^{\nu}{}_{\gamma\b\a}K^I_\nu.
\end{equation}
Taking these steps yields
\begin{eqnarray}\label{eq:cs-postulate}
    D_\mu J^{\mu I} =  s_K K_\nu^I\left(D_\mu T^{\mu\nu}+\inv{2}R^{\nu}_{\,\a\b\gamma}\mcS^{\a\b\gamma}\right) + s_K K^I_{[\a;\b]}\left(T^{[\a\b]}+\inv{2}D_\mu \mcS^{\mu\a\b}\right)\;.
\end{eqnarray}
The terms in parentheses must vanish independently, leading to the equations of motion in curved spacetime
\begin{subequations}
\begin{eqnarray}
    D_\mu T^{\mu\nu}&=&-\frac{1}{2}R^{\nu}{}_{\a\b\gamma}\mcS^{\a\b\gamma}\;,\label{eq:cs-emt}
    \\ 
    T^{[\a\b]}&=&-\frac{1}{2}D_\mu \mcS^{\mu\a\b}\;.\label{eq:cs-spin-dyn}
\end{eqnarray}
\label{eq:cs-spin}
\end{subequations}
These results align with those of Ref. \cite{HEHL197655}, where a variational approach was used assuming a nonvanishing torsion, whereas our derivation assumes a torsionless metric.

The equations \eqref{eq:cs-spin} are not covariant under the pseudo-gauge transformations \eqref{eq:psgt} if $\Xi^{\mu\nu\lambda\rho}\neq 0$. 
As demonstrated explicitly in Appendix \ref{app:curved}, the pseudo-gauge transformation must be generalized as 
\begin{equation}\label{eq:cs-psgt}
    T^{\mu\nu}{}' = T^{\mu\nu} + D_\lambda \Tilde{Z}^{\lambda\mu\nu}\;,\qquad S^{\lambda\mu\nu}{}' = S^{\lambda\mu\nu} - \Tilde{\Phi}^{\lambda\mu\nu} \;, 
\end{equation}
where 
\begin{equation}\label{eq:curved-Z}
    \Tilde{Z}^{\lambda\mu\nu} \equiv \inv{2}\left( \Tilde{\Phi}^{\lambda\mu\nu} - \Tilde{\Phi}^{\mu\lambda\nu} - \Tilde{\Phi}^{\nu\lambda\mu}\right)\;.
\end{equation}
Here, the definition of the superpotential is extended by adding a correction term,
\begin{equation}\label{eq:cs-superpotential}
    \Tilde{\Phi}^{\lambda\mu\nu} = \Phi^{\lambda\mu\nu} + \delta\Phi^{\lambda\mu\nu}\;,
\end{equation}
where the correction $\delta\Phi^{\lambda\mu\nu}$ maintains covariance of the equations of motion \eqref{eq:cs-spin}.
It is expressed as a sum of $N$ number of terms, where $N$ is arbitrary, as
\begin{eqnarray}
    \delta\Phi^{\lambda\mu\nu}= \sum_{n = 1}^{N} F_n^{\lambda\mu\nu} \qq{with} F_n^{\lambda\mu\nu} = \left(\prod_{k = 1}^{2n-1} D_{\r_k} \right)\Psi_n^{\mu\nu\lambda\r_{2n-1}\cdots\r_1}\;,
\end{eqnarray}
with $\Psi_n$ being a generic rank-$2(n+1)$ tensor, antisymmetric in adjacent pair of indices, such as $(\mu,\nu)$, and $(\lambda,\rho_{2n-1})$.

The term involving the tensor $\Xi^{\mu\nu\lambda\rho}$ in Eq.\ \eqref{eq:psgt} corresponds to $F_1^{\lambda\mu\nu}$ in our notation.
In flat spacetime, contributions from $\d \Phi^{\lambda\mu\nu}$ to transformations of the energy-momentum tensor vanish, as the covariant derivatives commute, reducing Eq.\ \eqref{eq:cs-psgt} to its standard form \eqref{eq:psgt}.
This observation explains why the extended form \eqref{eq:curved-Z} is not often reported in the literature.

Under the pseudo-gauge transformation \eqref{eq:cs-psgt}, the currents $J^{\mu I}$ transform as
\begin{equation}
    J^\mu{}^I{}' = J^\mu{}^I + s_K D_\lambda A^{\mu\lambda}{}^I\;, \qq{where} {A^{\mu\lambda}{}^I = \Tilde{Z}^{\lambda\mu\nu} K_\nu^I}\;,
\end{equation}
implying that, by Stokes’ theorem, the corresponding charges $Q^I$ remain invariant provided $A^{\mu\lambda}{}^I$ vanishes on the boundary.

The pseudo-gauge transformations were originally introduced to symmetrize the canonical energy-momentum tensor, derived using Noether’s theorem, making it compatible with the Einstein equation \cite{itzykson2012quantum},
\begin{equation}\label{eq:einstein}
     8\pi GT^{\mu\nu}_{\rm H} = G^{\mu\nu}\;,
\end{equation}
where $G$ is the gravitational constant, and $G^{\mu\nu} = R_{\mu\nu} - \tfrac{1}{2}R g_{\mu\nu}$ is the Einstein tensor.
The tensor $T^{\mu\nu}_{\rm H}$, known as the Hilbert energy-momentum tensor, arises naturally in the derivation of Einstein's field equation as the functional derivative of action $S$ with respect to the metric \cite{Poisson:2009pwt}
\begin{equation}\label{eq:hilbert}
    T^{\mu\nu}_{\rm H} = - \frac{2}{\sqrt{-g}}\fdv{S}{g_{\mu\nu}}\;.
\end{equation}
The Belinfante-Rosenfeld pseudo-gauge is commonly applied to symmetrize the energy-momentum tensor and eliminate the spin tensor \cite{belinfante1939spin,belinfante1940current,rosenfeld1940energy}.
The relation between the Belinfante-Rosenfeld energy-momentum tensor and an energy-momentum tensor in another pseudo-gauge is given by
\begin{equation}\label{eq:belifante}
    T^{\mu\nu}_{\rm B} = T^{\mu\nu} + \inv{2}D_\lambda \left(\mcS^{\lambda\mu\nu}-\mcS^{\mu\lambda\nu}-\mcS^{\nu\lambda\mu}\right)\;.
\end{equation}
However, the Belinfante-Rosenfeld tensor does not necessarily coincide with the Hilbert tensor \cite{Baker:2020eqs,Weldon:2023ztb}.
In cases where they do coincide, Eq.\ \eqref{eq:belifante} can be substituted into the Einstein equation \eqref{eq:einstein}, allowing it to be expressed in an arbitrary pseudo-gauge as
\begin{equation}
    8\pi GT^{(\mu\nu)} = G^{\mu\nu}+4\pi G D_\lambda\left(\mcS^{\mu\lambda\nu}+\mcS^{\nu\lambda\mu}\right)\;.
\end{equation}

\section{Semi-classical spin hydrodynamics}\label{sec:semi-classical}

As explained in the previous section, dissipative hydrodynamics is obtained by expanding hydrodynamic currents around a locally mapped reference (global) equilibrium state.
In addition to the classical hydrodynamic expansion, semi-classical spin hydrodynamics employs the so-called semi-classical expansion, where quantities are systematically expanded in $\hbar$ to account for quantum corrections.

This section begins by introducing key assumptions of semi-classical spin hydrodynamics.
Although these assumptions are inspired by spin hydrodynamics based on quantum kinetic theory \cite{Weickgenannt:2022zxs,Wagner:2024fry}, they are used to develop semi-classical spin hydrodynamics without directly relying on a microscopic theory.
Following these references, we truncate equations, rather than quantities, at first order in $\hbar$.
This first-order truncation has crucial implications, which will be clarified shortly.
After discussing the implications of these assumptions, we briefly review the ideal-spin hydrodynamics of Ref.\ \cite{Wagner:2024fhf}.

We emphasize that ``ideal-spin'' does not imply the absence of dissipation in the spin sector.
Rather, this term denotes an approximation where (1) the spin dynamics \revis{equation \eqref{eq:spin-dynamics-cartesian}} is closed, and (2) only contributions to the spin tensor that persist in global equilibrium are included.

\subsection{Key assumptions of semi-classical spin hydrodynamics}
As discussed in Appendix \ref{app:slow-rotation} and previously noted by Israel and Stewart in Ref.\ \cite{Israel:1979wp}, a fast-rotating reference equilibrium state introduces complications, such as spatial anisotropy at zeroth order. 
For example, the zeroth-order energy-momentum tensor deviates from the form \eqref{eq:perfect-fluid}.

To avoid these complications, we assume that the reference equilibrium state corresponds to a slowly rotating configuration, as defined in Appendix \ref{app:slow-rotation}.
These conditions ensure small equilibrium gradients, allowing the reference equilibrium state to remain approximately spatially isotropic.
Furthermore, we make the following assumptions learned from the quantum kinetic theory-based spin hydrodynamics:
\footnote{\revis{If assumption (A2), a reasonable assumption in heavy-ion physics, is relaxed, then the fluid might receive feedback from the spin dynamics even at order $\hbar$.
Therefore, it is reasonable to believe that there might be other physical systems where these assumptions do not hold.}}
\begin{itemize}
    \item [\textbf{(A1)}]\textbf{Symmetric energy-momentum tensor in equilibrium:}
    In global equilibrium, we choose a pseudo-gauge that makes the energy-momentum tensor symmetric while retaining the spin tensor:
    \begin{equation}
        T^{[\mu\nu]}_{\rm GTE} = 0\;,\qquad D_\lambda S^{\lambda\mu\nu}_{\rm GTE} = 0\;.
    \end{equation}
We apply this pseudo-gauge, known as de Groot-van Leeuwen-van Weert pseudo-gauge \cite{DeGroot:1980dk}, exclusively in global equilibrium.
   \item [\textbf{(A2)}] \textbf{Small polarization:} The spin part of angular momentum is significantly smaller than the orbital part, reflected in the rewritten form of Eq.\ \eqref{eq:tot-angular-momentum},
\begin{equation}
    J^{\lambda} = L^{\lambda} + \hbar S^{\lambda}\;,
\end{equation}
where $\hbar S^{\lambda} \equiv \mcS^{\lambda}$.
Thus, we also rewrite $\mcS^{\lambda\mu\nu} \equiv \hbar S^{\lambda\mu\nu}$ and express the spin dynamics \revis{equation \eqref{eq:spin-dynamics-cartesian}} as
\begin{equation}\label{eq:spin-dynamics}
   \hbar D_\lambda S^{\lambda\mu\nu} = 2T^{[\nu\mu]}\;.
\end{equation}
\item [\textbf{(A3)}] \textbf{No back-reaction from spin to fluid at first order in $\hbar$:} The quantum corrections to the energy-momentum tensor emerge only at second order in $\hbar$.
As a result, the antisymmetric part of the energy-momentum tensor decouples from Eq.\ \eqref{eq:em-cons} at the assumed first-order truncation,
\begin{equation}\label{eq:fluid-eom}
    D_\mu T^{(\mu\nu)} = \order{\hbar^2}\;.
\end{equation}
\end{itemize}
\subsection{Fluid sector}
The symmetric part of the energy-momentum tensor $T^{(\mu\nu)}$ appears only in Eq.\ \eqref{eq:em-cons}.
Therefore, up to the first order in $\hbar$, its components are purely \textit{fluid} fields.
For this reason, we refer to $T^{(\mu\nu)}$ as the fluid sector, whose components form the set $\{\varphi^H\}$.
Each quantity $\varphi^H$ represents an independent component, with $H=1\cdots N_f$ and $N_f$ being the number of independent components.

To identify these independent components, we first decompose $T^{(\mu\nu)}$ with respect to $u_\mu$ as \cite{Israel:1979wp,Kovtun:2012rj}:
\begin{equation}
	T^{(\mu\nu)} = \mcE u^\mu u^\nu - \mcP \Delta^{\mu\nu} + \mcQ^\mu u^\nu + \mcQ^\nu u^\mu + \mcT^{\mu\nu}\;,
\end{equation}
where 
\begin{equation}\label{eq:emt-comps-original}
	\mcE = u^\a u^\b T_{\a\b}\;,\qquad
	\mcP = -\frac{1}{3}\Delta_{\a\b} T^{\a\b}\;,\qquad
	\mcQ^\mu = \Delta^{\mu\a}u^\b  T_{\a\b}\;,\qquad
	\mcT^{\mu\nu} = \Delta^{\mu\nu}_{\alpha\beta} \, T^{\a\b}\;.
\end{equation}
To ascribe physical meanings to these components, the LTE mapping described in Sec.\ \ref{sec:geq-leq} must be applied.
The procedure is similar to IS-type dissipative hydrodynamic theories in the Landau frame, as it is carried out up to the first order in $\hbar$.
This implies that the \textit{semi-classical Landau frame} is defined as
\begin{equation}
    u_\mu T^{\mu\nu} = \varepsilon u^\nu + \order{\hbar^2}\;,
\end{equation}
and the semi-classical matching condition as 
\[
u^\a u^\b T_{\a\b} = \varepsilon + \order{\hbar^2}\;.
\]
Consequently, the components \eqref{eq:emt-comps-original} become
\begin{equation}\label{eq:emt-comps-landau}
	\mcE = \varepsilon + \order{\hbar^2}\;,\qquad
	\mcP = P + \Pi + \order{\hbar^2}\;,\qquad
	\mcQ^\mu = \order{\hbar^2}\;,\qquad
	\mcT^{\mu\nu} = \pi^{\mu\nu} + \order{\hbar^2}\;,
\end{equation}
where $\Pi$ is the bulk viscous pressure and $\pi^{\mu\nu}$ is the shear stress tensor.
Therefore, the set $\{\varphi^H\}$ contains $N_f=10$ independent components, $\{\varepsilon,\,u\;,\Pi\;,\pi^{\mu\nu}\}$, with $P$ being excluded using the equation of state (EOS) $P=P(\varepsilon)$.
The first-order truncation in $\hbar$ also keeps the spinless thermodynamic identities valid:
\begin{equation}\label{eq:thermo-id}
    \dd{\varepsilon} = T\dd{s} + \order{\hbar^2}\;,\qquad \dd{P} = s\dd{T} + \order{\hbar^2}\;,\qquad \varepsilon+P=Ts+ \order{\hbar^2}\;,
\end{equation}
where $s$ is the entropy density.

With definitions \eqref{eq:emt-comps-landau}, we project Eq.\ \eqref{eq:fluid-eom} into longitudinal and orthogonal directions relative to $u_\mu$, yielding
\begin{subequations}\label{eq:fluid-eom-decomp}
\begin{eqnarray}
    \dot{\beta} &=& -\frac{1}{\pdv*{\varepsilon}{\b}}\left[(\varepsilon+P+\Pi)\theta - \pi^{\a\b}\sigma_{\a\b}\right]\;,\label{eq:bdot}
    \\
   a_\mu &=& -\frac{1}{\varepsilon+P+\Pi}\left[(\varepsilon+P)\nabla_\mu\ln{\beta}-\Delta_{\mu\a}D_\nu\pi^{\nu\a}-\nabla_\mu \Pi\right]\;,\label{eq:gradb}
\end{eqnarray}
\end{subequations}
where $\beta=1/T$ is the inverse temperature.

To close Eq.\ \eqref{eq:em-cons}, equations of motion for $\Pi$ and $\pi^{\mu\nu}$ are required in addition to the EOS.
For simplicity, we state the equation of motion for the bulk viscous pressure as
\begin{equation}
    \tau_{\Pi}\dot{\Pi} + \Pi = -\zeta\theta + \cdots\;,
\end{equation}
where $\tau_{\Pi}$ is the bulk relaxation time and $\zeta$ is the bulk viscosity coefficient. 
Additional terms can be found, e.g., in Refs.\ \cite{Israel:1979wp} and \cite{Denicol:2012cn}.
For the shear stress tensor, the DNMR equation of motion is assumed \cite{Denicol:2012cn}:
  \begin{align}\label{eq:shear-eom}
\tau_{\pi}  \dot{\pi}^{\langle \mu \nu \rangle}    + \pi^{\mu \nu} &= 2 \eta \sigma^{\mu \nu} 
+ 2 \tau_{\pi} \pi_{\lambda}^{\langle \mu} \fvort^{\nu \rangle \lambda}  
- \delta_{\pi \pi} \pi^{\mu \nu} \theta  
 - \tau_{\pi \pi} \pi^{\lambda \langle \mu} \sigma^{\nu \rangle}_{\lambda} + \cdots\;,
\end{align}
where $\tau_{\pi}$ is the shear relaxation time, $\eta$ is the shear-viscosity coefficient, and $\delta_{\pi \pi}$ and $\tau_{\pi \pi}$ are second-order transport coefficients.
Terms not used in this work are omitted.

\subsection{The spin sector and ideal-spin approximation}
We now turn to the spin dynamics equation \eqref{eq:spin-dynamics}, which, along with assumption (A3), indicates that the spin tensor $S^{\lambda\mu\nu}$ is first order in $\hbar$. 
The independent components in $S^{\lambda\mu\nu}$ form the set $\{\psi^L\}$, where each $\psi^L$ represents an independent component, and $L=1,\dots, N_s$, with $N_s$ denoting the number of independent components in $S^{\lambda\mu\nu}$. 
Note that the maximum number of independent components in $S^{\lambda\mu\nu}$ is $24$.

The spin tensor $S^{\lambda\mu\nu}$ can be decomposed as 
\begin{equation}
    S^{\lambda\mu\nu} = S^{\lambda\mu\nu}_{(0)} + \d S^{\lambda\mu\nu}\;,
\end{equation}
where $S^{\lambda\mu\nu}_{(0)}$ persists in global equilibrium, while $\d S^{\lambda\mu\nu}$ does not \cite{Wagner:2024fhf}.
Analogously to the perfect fluid, whose energy-momentum tensor is identical \textit{in form} to its global equilibrium counterpart, the ideal-spin approximation corresponds to neglecting $\d S^{\lambda\mu\nu}$.

We expect the spin dynamics equation \eqref{eq:spin-dynamics} to be closed and trivial in global equilibrium, implying that the ideal-spin tensor $S^{\lambda\mu\nu}_{(0)}$ contains $6$ degrees of freedom.
Therefore, out of equilibrium,  $S^{\lambda\mu\nu}_{(0)}$ is linear in the so-called spin potential tensor $\Omega_{\mu\nu}$, an antisymmetric tensor that becomes equal to the thermal vorticity $\varpi_{\mu\nu}$ in global equilibrium.
As with $\varpi_{\mu\nu}$, the spin potential can be decomposed into electric and magnetic parts,
    \begin{equation}
\Omega^{\mu\nu}=2u^{[\mu}\kappa^{\nu]}+\epsilon^{\mu\nu\alpha\beta}u_\alpha \omega_{\beta}\;,
\label{eq:decomp_Omega}
\end{equation}
where 
\begin{equation}
\kappa^\mu \equiv -\Omega^{\mu\nu}u_\nu \;,\qquad \omega^\mu \equiv \frac12\epsilon^{\mu\nu\alpha\beta}u_\nu \Omega_{\alpha\beta}\;.
\end{equation}
The ideal-spin tensor then takes the following general form:
\begin{align}
S^{\lambda\mu\nu}_{(0)}=A  u^\lambda \Omega^{\mu\nu}+2B u^\lambda u_\alpha \Omega^{\alpha[\mu} u^{\nu]} + 2C u^\lambda \Omega^{\alpha[\mu}\Delta^{\nu]}{}_\alpha
+2Du_\alpha \Omega^{\alpha[\mu}\Delta^{\nu]\lambda}+2E\Delta^\lambda{}_\alpha \Omega^{\alpha[\mu}u^{\nu]}
\;,\label{eq:decomp_S}
\end{align} 
where the coefficients $\{A,\, B,\, C,\, D,\, E\}$ are functions of the fluid variables, first-order in $\hbar$.

Next, we turn to the right-hand side of Eq.\ \eqref{eq:spin-dynamics}, i.e., the antisymmetric part of the energy-momentum tensor, which takes the following form:
\begin{align}
    T^{[\mu\nu]}&=-\hbar^2\Gamma^{(\kappa)} u^{[\mu} \left(\kappa^{\nu]}{+}  \varpi^{\nu]\alpha}u_\alpha \right)  + \inv{2}\hbar^2\Gamma^{(\omega)} \epsilon^{\mu\nu\rho\sigma}u_\rho \left(\omega_\sigma+\beta \kvort_\sigma \right) + \hbar^2\Gamma^{(a)} u^{[\mu} \left( \beta a^{\nu]}+\nabla^{\nu]}\beta\right)\;,
    \label{eq:T_A}
\end{align}
where the coefficients $\Gamma^{(\kappa)}$, $\Gamma^{(\omega)}$, and $\Gamma^{(a)}$ are functions of the fluid variables.
For an uncharged perfect fluid, the term multiplied with $\Gamma^{(a)}$ vanishes and, therefore, arises solely from dissipative contributions in the fluid sector.

In global equilibrium, all terms in Eq.\ \eqref{eq:T_A} vanish as is required by assumption (A1).
Enforcing $D_\lambda S^{\lambda\mu\nu}=0$ in global equilibrium, on the other hand, gives rise to the following constraint:
\begin{equation}\label{eq:coeffs_constraint}
     B-C-D-\beta\pdv{E}{\beta} = 0\;.
\end{equation}

It is useful to decompose the spin dynamics equation \eqref{eq:spin-dynamics} to write it in terms of the magnetic and electric components of the spin potential.
For this purpose, we contract Eq.\ \eqref{eq:spin-dynamics} with $u_\mu$ and $\inv{2}\epsilon_{\mu\nu\a\b}u^b$ and then use Eq.\ \eqref{eq:T_A}.
This leads to
\begin{eqnarray}\label{eq:ideal-spin-kappa}
     (A-B-C) \dot{\kappa}^{\langle\mu\rangle} &=& - \left[\left(\pdv{A}{\beta}-\pdv{B}{\beta}-\pdv{C}{\beta}\right)\dot{\beta}+\left(A-B-C+\frac{2}{3}D\right)\theta\right]\kappa^\mu 
    +  D \left(\fvort^{\mu\lambda}+\sigma^{\mu\lambda}\right)\kappa_\lambda 
    \n
    &&+\epsilon^{\mu\nu\rho\sigma}u_\nu \left[(E+2C-A)a_\rho \omega_\sigma + E \,\nabla_\rho \omega_\sigma + \left(\pdv{E}{\b} \nabla_\rho \beta\right) \omega_\sigma\right]
    \n
    &&+ \hbar \Gamma^{(\kappa)}\left(\kappa^\mu + \varpi^{\mu\nu}u_\nu\right) -\hbar\Gamma^{(a)}\left(\nabla^\mu\beta + \beta a^\mu\right)\;,
\end{eqnarray}
and
\begin{eqnarray}\label{eq:ideal-spin-omega}
   (A-2C) \dot{\omega}^{\langle\mu\rangle} &=& - \left[\left(\pdv{A}{\beta}-2\pdv{C}{\beta}\right)\dot{\beta}+\left(A-2C-\frac{2}{3}E\right)\theta\right]\omega^\mu 
    -  E \left(\fvort^{\mu\lambda}+\sigma^{\mu\lambda}\right)\omega_\lambda 
    \n
    &&+\epsilon^{\mu\nu\rho\sigma}u_\nu \left[(B+C-A-D)a_\rho \kappa_\sigma + D \,\nabla_\rho \kappa_\sigma + \left(\pdv{D}{\b} \nabla_\rho \beta\right) \kappa_\sigma\right]
    \n
    &&-\hbar \Gamma^{(\omega)}\left(\omega^\mu + \beta \kvort^\mu \right)\;.
\end{eqnarray}
In these equations, the quantities $\dot{\beta}$ and $\theta$, on one hand, and $a_\mu$ and $\nabla_\mu\beta$, on the other hand, are not independent but are related through Eqs.\ \eqref{eq:bdot} and \eqref{eq:gradb}, respectively.

Equations \eqref{eq:ideal-spin-kappa} and \eqref{eq:ideal-spin-omega} are relaxation-type equations and, therefore, possess a dissipative nature.
However, the associated entropy production is second order in $\hbar$ and is neglected.
Inserting the following coefficients derived using quantum kinetic theory \cite{Wagner:2024fhf} into Eqs.\ \eqref{eq:ideal-spin-kappa} and \eqref{eq:ideal-spin-omega}, the equations of motion of Ref.\ \cite{Wagner:2024fry} are recovered in the ideal-spin approximation:
\begin{equation}\label{eq:kinetic-coefficients}
   A=\frac{\hbar T^2}{4m^2}\pdv{T}\left(\varepsilon-3P\right)\;\qquad B=\frac{\hbar T^2}{4m^2} \pdv{\varepsilon}{T}\;\qquad C=D=E=-\frac{\hbar T^2}{4m^2}\pdv{P}{T}\;.
\end{equation}

At first order in $\hbar$, spin does not back-react to the fluid, allowing solutions to the fluid equations of motion to serve as input to the spin dynamics equation \eqref{eq:spin-dynamics}.
In this work, we assume two simple solutions to the fluid's equation of motion---linear waves and conformal Bjorken flow---to explore fluid-spin dynamics.

\section{Linearized spin hydrodynamics}\label{sec:linear-spin-hydro}

In this section, we explore semi-classical spin hydrodynamics in the linear regime.
For this purpose, we perturb fields $X$, which belong to the set of fluid fields ${\varphi^H}$ or spin fields ${\psi^L}$, around a homogeneous equilibrium configuration as
\begin{equation}\label{eq:x-def}
    X = X_\eq + \d X\;.
\end{equation}
The subscript ``$\eq$'' denotes equilibrium and the index $A$ labels the independent components of the fields.
In a homogeneous equilibrium, $X_\eq$ are constant in space and time.

Subsequently, we transform the quantities into the momentum space using the Fourier transform defined as
\begin{equation}\label{eq:fourier}
    \d X(x) = \int\frac{\dd[4]{k}}{(2\pi\hbar)^4}e^{ik\cdot x/\hbar}\d X(k)\;,
\end{equation}
where the factor $\hbar$ in the exponential, and the integration measure, is introduced for dimension consistency and not semi-classical power counting.
For simplicity, we use the same symbol $\d X$ to represent quantities in both $x$ space and $k$ space, as the context will make it clear which space is being referred to.

The set $\{\varphi^H\}$ comprises all relevant fluid fields truncated at first order in $\hbar$, which may include not only the ones introduced in the previous section but also multiple charge densities and charge diffusion currents \cite{Gavassino:2023qnw}.
The evolution of these variables is governed by the truncated energy-momentum conservation \eqref{eq:fluid-eom}, conservation of charge currents, and equations of motion for the dissipative fluxes.
Together, these equations form the set referred to as fluid equations of motion.

Upon inserting Eq.\ \eqref{eq:x-def}, with $X=\varphi^H$, into the fluid equations of motion, keeping only terms first order in perturbations, and subsequently performing the Fourier transform \eqref{eq:fourier}, the equations yield a set of linear algebraic equations.
This set is expressed in matrix form as
\begin{equation}\label{eq:fluid-collective}
    M^{HK}_{\rm f} \d\varphi^{K} = 0\;,
\end{equation}
where the Einstein summation convention is assumed for the index $K=1\cdots N_{\rm f}$.
Here, the components of $M_{\rm f}$ are spacetime-independent quantities that depend on the momentum $k_\mu$, as well as fluid variables and transport coefficients in equilibrium.
The right-hand side of this equation is zero because the fluid variables are perturbed around a homogeneous equilibrium state.

Now, let us turn to the spin sector.
As mentioned in the previous section, $S^{\lambda\mu\nu}$ can have up to 18 additional degrees of freedom beyond the 6 ones associated with the spin potential $\Omega^{\mu\nu}$.
Consequently, the form \eqref{eq:decomp_S} is not general, allowing for additional \textit{non-ideal} terms \cite{Weickgenannt:2022zxs, Wagner:2024fry}.
The presence of non-ideal spin fluxes implies that Eq.\ \eqref{eq:spin-dynamics} is no longer closed, necessitating additional equations of motion.
These equations incorporate contributions from both spin and fluid variables.
After linearization via Eq.\ \eqref{eq:x-def} and Fourier transformation via Eq.\ \eqref{eq:fourier}, these equations take a matrix form similar to fluid equations of motion
\begin{equation}\label{eq:spin-collective}
  M^{XY}_{\rm s} \d\psi^{Y} +  M^{XY}_{\rm fs} \d\varphi^{Y} = 0\;,
\end{equation}
where Einstein summation convention is assumed for the index $Y=1\cdots N_{\rm s}$.
Here, the fluid-spin coupling matrix $M_{\rm fs}$ arises from the coefficients of the source terms in Eq.\ \eqref{eq:spin-dynamics}, i.e., $T^{[\mu\nu]}$, as well as the equations of motion for the non-ideal spin fluxes.
The components of both $M_{\rm s}$ and $M_{\rm fs}$ are, like those of $M_{\rm f}$, spacetime-independent and depend on momentum and equilibrium quantities. 

Equations \eqref{eq:fluid-collective} and \eqref{eq:spin-collective} can be combined into a single $N\times N$ matrix equation, where $N=N_{\rm f}+N_{\rm s}$:
\begin{eqnarray}\label{eq:spin-fluid-collective}
    \left(\begin{array}{cc}
       M_{\rm f}  & 0 \\
       M_{\rm fs}  &  M_{\rm s} 
    \end{array}\right)\left(
    \begin{array}{c}
         \d\varphi  \\
         \d\psi
    \end{array}\right)
    &=& \vb{0}\;.
\end{eqnarray}
This system of homogeneous linear equations is solvable only if $\det M = 0$, which yields the \textit{characteristic equation} \cite{Hiscock:1985zz, Shokri:2023rpp}.
The characteristic equation is a polynomial in terms of $\omega_t = k\cdot t$, where $t_\mu$ denotes the frame vector, and $\omega_t$ is the frequency of the plane waves in the chosen Lorentz frame.
Solving the characteristic equation leads to dispersion relations, which determine $\omega_t$ in terms of spatial components of $k_\mu$.
The dispersion relations identify different wave propagation modes and provide a criterion for the linear stability of the hydrodynamic theory.
In our convention, the theory is linearly stable if $\Im \omega_t \geq 0$ for any arbitrary choice of $t_\mu$ \cite{Hiscock:1985zz}.

We now prove a crucial result regarding the linear wave analysis in semi-classical spin hydrodynamics.
First, we observe that the determinant of the following matrix is $1$, so multiplying it with the matrix Eq.\ \eqref{eq:spin-fluid-collective} does not change the determinant:
\begin{equation}
    \left(\begin{array}{cc}
        1 & 0 \\
        -M^{-1}_{\rm s}M_{\rm fs} & 1
    \end{array}\right)\;.
\end{equation}
Thus, we find
\begin{equation}
    \det \left(\begin{array}{cc}
       M_{\rm f}  & 0 \\
       M_{\rm fs}  &  M_{\rm s} 
    \end{array}\right) = \det\left(\begin{array}{cc}
       M_{\rm f}  & 0 \\
      0 &  M_{\rm s} 
    \end{array}\right) = \det( M_{\rm f}) \det( M_{\rm s} )\;.
\end{equation}
This shows that the fluid-spin coupling matrix $M_{\rm fs}$ decouples from the characteristic equation, which separates into independent fluid and spin parts:
\begin{equation}
    \det( M_{\rm f}) = 0 \;,\qquad \det( M_{\rm s} ) = 0\;.
\end{equation}
In other words, we have demonstrated that if spin does not back-react to the fluid dynamics, the linear characteristic equation determining the spin dispersion relations decouples from the fluid modes.
A concrete example illustrating this general result is presented in Sec.\ \ref{sec:example}.

This result shows that, at the first order in $\hbar$, the spin sector's linear wave analysis can be performed assuming the fluid sector remains in equilibrium, without loss of generality.
However, the resulting dispersion relations are inherently first order in $\hbar$ and, making the linear stability criterion an approximate one due to the first-order truncation in the semi-classical spin hydrodynamics.
In the next section, we demonstrate that this approximate nature also applies to the Gibbs stability criterion.

\section{Gibbs stability in semi-classical hydrodynamics}\label{sec:entropy}

In this section, we review the Gibbs stability criterion \cite{Hiscock:1983zz,Gavassino:2021cli} and demonstrate its limitations in semi-classical hydrodynamics.
These limitations are analogous to those encountered in the previous section and arise from truncations of the equations at first order in $\hbar$.

To illustrate this, we consider a closed system comprising a body and its surrounding environment.
As noted in Sec.\ \ref{sec:geq-leq}, the body reaches equilibrium with the environment when the total entropy is maximized, subject to constraints imposed by the conserved charges $\{Q^I\}$, as introduced in Sec.\ \ref{sec:spin-hydro}.

We assume the interaction between the body and the environment is sufficiently weak so that the entropies of the body $S$ and the environment $S_E$ depend only on their respective charges $Q^I$ and $Q^I_E$ \cite{Gavassino:2023qnw}.
This assumption implies that, in equilibrium, the density operators for both the body and the environment are defined using the same set of Lagrange multipliers $\lambda_I$, as given in Eq.\ \eqref{eq:lambda-coeffs}:
\begin{equation}\label{eq:geq_rho}
    \hat{\rho} = \inv{Z}e^{-\lambda_I \hat{Q}^I}\;,
\end{equation}
where $\hat{Q}^I$ are operators with expectation values $Q^I=\Tr(\hat{\rho}\hat{Q}^I)$, and $Z = \Tr[\exp(-\lambda_I \hat{Q}^I)]$ is the partition function \cite{Balian2007}.

Now, we perturb the body's state by varying its density operator $\hat{\rho}$ to a neighboring normalized density operator, $\hat{\rho}'=\hat{\rho}+\d\hat{\rho}$, where $\d\hat{\rho}$ represents a small deviation.
This perturbation shifts the body to a state that is slightly out of equilibrium, with the charges $Q^I{}'$ in this new state given by $\Tr(\hat{\rho}'\hat{Q}^I)$.
According to the Bogoliubov inequality (see, e.g., Ref.\ \cite{Balian2007} Ch.\ 4.4.2),
\begin{equation}
    S[\hat{\rho}'] - \lambda^\star_I \Tr(\hat{\rho}'\hat{Q}^I) < \ln{Z}\;.
\end{equation}
Thus, the function $\Phi \equiv S - \lambda^\star_I Q^I$ is maximized in equilibrium,
\begin{equation}\label{eq:phi_bound}
    \Phi \leq \ln{Z}\;.
\end{equation}

By using Eqs.\ \eqref{eq:cons_current} and \eqref{eq:cons_charges}, and expressing $\Phi$ as a surface integral over an arbitrary Cauchy hypersurface $\Sigma$, $\Phi = \int_{\Sigma}\dd{\Sigma}_\mu\phi^\mu$, we identify the \textit{Gavassino} current of spin hydrodynamics as
\begin{equation}\label{eq:phi_vector}
    \phi^\mu = S^\mu + \xi^\star N^\mu - T^{\mu\nu}\beta^\star_\nu + \frac{1}{2} \mcS^{\mu\a\b}\varpi^\star_{\a\b}\;.
\end{equation}
Here, in contrast to the other sections of this work, we have assumed a charged fluid that has a single charge density current $N^\mu$, which satisfies the charge density current conservation $N^\mu_{;\mu}=0$.
The constant $\xi^\star$ is defined as
\begin{equation}
    \xi^\star = -\pdv{S}{N}\Big|_{\rm GTE}\;.
\end{equation}
Using Eqs.\ \eqref{eq:em-cons} and \eqref{eq:spin-dynamics}, the charge density current conservation, and the second law of thermodynamics, we find
\begin{equation}
    D_\mu\phi^\mu = D_\mu S^\mu \geq 0\;.
\end{equation}

Perturbations of the spin and fluid fields, $X \in  \{\varphi^H\} \cup \{\psi^L\}$, can be formally expressed in terms of their first-order derivative with respect to a common perturbation variable $\lambda$ as
\begin{equation}
\d X = \dv{X}{\lambda} \d\lambda + \order{\d\lambda^2} \;.
\end{equation}

By taking the first-order derivative of the vector $\phi^\mu$ and setting it to zero, one finds the stationary points, which without loss of generality, we assume occur at $\lambda = 0$.
To determine whether a stationary point corresponds to a true equilibrium state, the second-order derivative is examined to obtain necessary equilibrium conditions (see, e.g., Ref.\ \cite{Gavassino:2023qnw} and references therein).
More precisely, one computes the so-called information current, defined as 
$$
E^\mu \equiv - \inv{2} \dv[2]{\phi^\mu}{\lambda} \;,
$$ 
and identifies the conditions that ensure it is future-directed and non-spacelike.
These conditions constitute the thermodynamic stability criteria for the corresponding theory.

To find the stationary point and identify the stability conditions, the constitutive relations for $T^{\mu\nu}$, $N^\mu$, $S^\mu$, and $\mcS^{\mu\a\b}$ are required.
At present, these are known only up to the first order in $\hbar$.
Consequently, $\phi^\mu$ defined in Eq.\ \eqref{eq:phi_vector}, must also be truncated at first order in $\hbar$ 
\begin{equation}\label{eq:phi_vector-hbar}
    \phi^\mu = S^\mu_{\rm f} + \xi^\star N^\mu_{\rm f} - T^{(\mu\nu)}_{\rm f}\beta^\star_\nu + \order{\hbar^2}\;.
\end{equation}
The stationary point and the information current derived from this truncated vector are identical to those obtained in the dissipative hydrodynamic theory without spin, corresponding to the theory's fluid sector.
As a result, only the stability of the fluid sector can be assessed when the semi-classical expansion is truncated at the first order.

Inserting the semi-classical relations for $\mcS^{\mu\a\b}$ and $T^{[\mu\nu]}$ into Eq.\ \eqref{eq:phi_vector} is not consistent, as it neglects the second-order quantum corrections to $S^\mu$, $N^\mu$, and $T^{(\mu\nu)}$.
Doing so would incorrectly imply that, in equilibrium, $\mcS^{\lambda\mu\nu}$ and $\varpi_{\mu\nu}$ vanish.

This confirms our earlier observation that the equilibrium gradients of reference equilibrium necessitate a revision of the fluid sector, accounting for the inherent anisotropy in equilibrium.
As mentioned above, we defer this discussion to future work.

In summary, the linear stability conditions of semi-classical hydrodynamics are valid only up to the first order in $\hbar$.

\section{An example: Linear ideal-spin hydrodynamics with shear viscosity}\label{sec:example}
In this section, we demonstrate the results of Sec.\ \ref{sec:linear-spin-hydro} using the example of linear ideal-spin hydrodynamics. 
This example incorporates a nonzero shear stress tensor while assuming a vanishing bulk viscous pressure.
We choose the independent fluid and spin fields as $\{\beta, u^\mu, \pi^{\mu\nu}, \kappa^\mu, \omega^\mu\}$, perturbing them as described in Sec.\ \ref{sec:linear-spin-hydro} to determine the dispersion relations for the fluid and spin waves.

For an uncharged fluid, it is convenient to assume the EOS in the form $\varepsilon = \varepsilon(P)$, which allows one to express the perturbation of the energy density as $\d\varepsilon = (\pdv*{\varepsilon}{P})\d P$.
Then, using the thermodynamic identities \eqref{eq:thermo-id}, we can relate the pressure perturbation to $\d\beta$ as $\d P = - \enp_\eq \d\beta / \beta_\eq$, where $\enp_\eq \equiv \varepsilon_\eq + P_\eq$ is the enthalpy density in equilibrium.

To handle the independent components of vector and tensor quantities, as well as their corresponding equations, in a covariant manner, we follow the method introduced in Ref.\ \cite{Brito:2020nou}.
First, we decompose the momentum $k^\mu$ as 
\begin{equation}\label{eq:k-decomp}
    k^\mu = \freq u_\eq^\mu + \ktsup{\mu} \;.
\end{equation}
where $\ktsup{\mu} \equiv \Delta^{\mu\nu}_\eq k_\nu$, with $\Delta^{\mu\nu}_\eq\equiv g^{\mu\nu} - u^\mu_\eq u^\nu_\eq$.
Here, $\freq$ represents the frequency in the \textit{equilibrium} local rest frame, where $u^\mu_\eq = (1,\vb{0})$.
In the \elrf, the spatial part of $k^\mu$ has a modulus given by $\kt \equiv \sqrt{-\ktsup{\mu}\ktsub{\mu}}$.

Using the decomposition \eqref{eq:k-decomp}, the spacetime is split into the timelike direction defined by $u^\mu_\eq$, the spacelike direction along $\ktsup{\mu}$, and the two-dimensional plane orthogonal to both $u^\mu_\eq$ and $\ktsup{\mu}$.
To facilitate the decomposition, we define the covariant projector $\projkt^{\mu\nu}$, which projects any vector onto the plane orthogonal to $u^\mu$ and $\ktsup{\mu}$,
\begin{equation}\label{eq:kt-projector}
    \projkt^{\mu\nu} \equiv g^{\mu\nu} - u^\mu_\eq u^\nu_\eq + \frac{1}{\ktsup{2}}\ktsup{\mu} \ktsup{\nu}\;.
\end{equation}
By inserting the momentum decomposition \eqref{eq:k-decomp} into the Fourier transform \eqref{eq:fourier}, we realize that the equations of motion in momentum space are obtained through the following replacements:
\begin{equation}\label{eq:fourier-rep}
    \dot{\d X} \to \frac{i}{\hbar}\freq \d X\;,\qquad \nabla^\mu \d X \to \frac{i}{\hbar} \ktsup{\mu} \d X\;,
\end{equation}
where $\d\dot{ X} \equiv u^\mu_\eq \partial_\mu \d X$ and $\nabla^\mu \d X = \Delta^{\mu\nu}_\eq \partial_\nu \d X$.

Using these replacements, the equations of motion are expressed in matrix form \eqref{eq:spin-fluid-collective}, and $\det M = 0$ determines the characteristic equation.
The roots of the characteristic equation, $\omega_a = \omega_a(\kt)$, where $a$ denotes the mode index, correspond to the frequencies of different modes in the \elrf.

Vector perturbed quantities are also decomposed into the directions specified here. 
In linear ideal-spin hydrodynamics, there are three independent vector degrees of freedom: $\{u^\mu,\,  \kappa^\mu,\, \omega^\mu\}$.
In a homogeneous equilibrium configuration, $u^\mu_\eq$ is constant, the equilibrium thermal vorticity $\varpi_{\mu\nu}$ vanishes, and, consequently, the equilibrium spin potential is also zero.
Thus, using $\kappa^\mu_\eq = \omega^\mu_\eq = 0$ and $u^\mu u_\mu =1$, we find 
\begin{equation}
    u^\mu_\eq \d u_\mu = u^\mu_\eq \d \kappa_\mu = u^\mu_\eq \d\omega_\mu = 0\;.
\end{equation}
As a result, the transverse vectors $\{\d u^\mu,\,  \d\kappa^\mu,\, \d\omega^\mu\}$ are decomposed into their independent components as
\begin{equation}\label{eq:generic-transverse-vector}
    V^\mu = V_\kt \frac{\ktsup{\mu}}{\kt} + V^\mu_\perp\;,
\end{equation}
where $V_\kt \equiv -\ktsub{\mu} V^\mu$ and $V^\mu_\perp \equiv \projkt^{\mu\nu} V_\nu$, with $V^\mu \in \{\d u^\mu,\, \d \kappa^\mu,\, \d\omega^\mu\}$.

Due to the presence of the Levi-Civita tensor in Eqs.\ \eqref{eq:ideal-spin-kappa} and \eqref{eq:ideal-spin-omega}, it is useful to introduce the following notation:
\begin{equation}\label{eq:generic-cross}
    V^\mu_\chi \equiv \inv{\kt}\epsilon^{\mu\nu\a\b}u^0_\nu\ktsub{\a} V_{\perp\b}\;,
\end{equation}
where, in the \elrf{}, this is proportional to $\va{\nabla}\times \vb{V}_\perp$.
This relation transforms an axial vector into a vector and vice versa.
Note that $\d\omega^\mu$ is an axial vector, while $\d\kappa^\mu$ is a vector.

The remaining rotational symmetry around $\ktsup{\mu}$ in the two-dimensional plane orthogonal to $u^\mu_\eq$ and $\ktsup{\mu}$ gives us the freedom to choose either the vector pair ${\d\kappa^\mu_\perp, \d\omega^\mu_\chi}$ or the axial vector pair ${\d\kappa^\mu_\chi, \d\omega^\mu_\perp}$ as the independent degrees of freedom.
In the following, we adopt the latter pair, which requires transforming all other vectors into their corresponding axial vector forms.

The only independent tensor quantity is the shear-stress tensor $\d \pi^{\mu\nu}$, which is decomposed as follows:
\begin{equation} \label{eq:shear-decomp}
    \d \pi^{\mu\nu} =  \d \pi_\kt\frac{1}{\ktsup{2}}\ktsup{\mu}\ktsup{\nu}+\inv{2}\d\pi_\kt\projkt^{\mu\nu}+\left(\d \pi^\mu_\perp \frac{\ktsup{\nu}}{\kt}+\d \pi^\nu_\perp \frac{\ktsup{\mu}}{\kt}\right)+ \d \pi^{\mu\nu}_\perp\;,
\end{equation}
where the components are defined as
\begin{equation}\label{eq:dpi-components}
     \d \pi_\kt \equiv \inv{\ktsup{2}}\ktsub{\mu}\ktsub{\nu} \d\pi^{\mu\nu} \;,\qquad
    \d\pi_\perp^\mu \equiv - \frac{\ktsup{\nu}}{\kt} \projkt^{\mu\rho}\d\pi_{\nu\rho}\;,\qquad
    \d\pi^{\mu\nu}_\perp = \projkt^{\mu\nu}_{\a\b} \d\pi^{\a\b}\;.
\end{equation}
The traceless symmetric projector of rank-4 in the plane orthogonal to $u^\mu_\eq$ and $\ktsup{\mu}$ is given by
\begin{equation}
    \projkt^{\mu\nu}_{\a\b} \equiv \frac{1}{2}\left(\projkt^{\mu}_{\alpha}\projkt^{\nu}_{\beta}+\projkt^{\mu}_{\beta}\projkt^{\nu}_{\alpha}\right)-\frac{1}{2}\projkt^{\mu\nu}\projkt_{\alpha\beta}\;.
\end{equation}

\subsection{Fluid sector linear equations of motion}

First, we consider the fluid equations of motion, i.e., Eqs.\ \eqref{eq:bdot}, \eqref{eq:gradb}, and \eqref{eq:shear-eom}.
A detailed treatment of these equations can be found for example in Refs.\ \cite{Brito:2020nou,Sammet:2023bfo}; thus, we briefly summarize the steps required to transform them into a system of algebraic equations in momentum space.
Starting from the linearized form of Eqs.\ \eqref{eq:bdot} and \eqref{eq:gradb}, we apply the replacements \eqref{eq:fourier-rep} to bring them to momentum space.
Subsequently, the linearized and transomed form of Eq.\ \eqref{eq:gradb} is contracted with $\ktsub{\mu}$ and $\projkt_{\mu\nu}$.
As a result, Eqs.\ \eqref{eq:bdot} and \eqref{eq:gradb} lead to the following system of equations: 
\begin{subequations}
    \label{eq:em-conservation-k}
    \begin{eqnarray}
         && \freq\left(\pdv{\varepsilon}{P}\right) \frac{\d\b}{\b_\eq} +\kt \d u_\kt = 0\;,\label{eq:em-cons-u}\\
         && \freq\, \d u_\kt +\kt\frac{\d\b}{\b_\eq}  -  \frac{\kt}{\enp_\eq}\d\pi_\kt = 0\;,\label{eq:em-cons-kt}\\
         && \freq\,\d u^{\nu}_{\perp} -\frac{\kt}{\enp_\eq}\d\pi^{\nu}_\perp = 0\;,\label{eq:em-cons-perp}
    \end{eqnarray}
\end{subequations}
where Eqs.\ \eqref{eq:generic-transverse-vector} and \eqref{eq:dpi-components} have been used.

To align with our choice of axial vector degrees of freedom specified earlier, we express Eq.\ \eqref{eq:em-cons-perp} in terms of the axial vectors $\d u^\mu_\chi$ and $\d\pi^\mu_\chi$, instead of the vectors $\d u^\mu_\perp$ and $\d\pi^\mu_\perp$.
To achieve this, we contract this equation with $\epsilon^{\mu}{}_{\nu\rho\sigma}u^\rho_\eq\ktsup{\sigma}/\kt$ and employ Eq.\ \eqref{eq:generic-cross}, giving rise to
\begin{equation}
    \freq\d u^{\mu}_{\chi} -\frac{\kt}{\enp_\eq}\d\pi^{\mu}_\chi = 0\;.\label{eq:em-cons-cross}
\end{equation}

Now, we turn to the equation of motion for the shear stress tensor \eqref{eq:shear-eom}.
To linearize this equation, we first note that the shear tensor $\sigma^{\mu\nu}$ vanishes in equilibrium, and therefore 
\begin{equation}\label{eq:dsigma}
\d \sigma^{\mu\nu} = \left(\Delta^{\mu\nu}_{\a\b}\right)_\eq \nabla^\a \d u^\b\;,
\end{equation}
where
\begin{equation}
    \left(\Delta^{\mu\nu}_{\a\b}\right)_\eq \equiv \frac{1}{2}\left(\Delta^{\mu}_{\alpha \eq}\Delta^{\nu}_{\beta \eq}+\Delta^{\mu}_{\beta \eq}\Delta^{\nu}_{\alpha\eq }\right)-\frac{1}{3}\Delta^{\mu\nu}_\eq\Delta_{\alpha\beta \eq}\;.
\end{equation}
Using the replacements \eqref{eq:fourier-rep} and the decomposition \eqref{eq:generic-transverse-vector} in Eq.\ \eqref{eq:dsigma} leads to the following relation for $\d \sigma^{\mu\nu}$ in momentum space:
\begin{equation}\label{eq:dsigma-decomp}
  \d \sigma^{\mu\nu} = \frac{i}{\hbar}\left[\frac{2}{3}\frac{\d u_\kt}{\kt}\ktsup{\mu}\ktsup{\nu}+\frac{1}{2}\left(\d u^\mu_\perp \frac{\ktsup{\nu}}{\kt}+\d u^\nu_\perp \frac{\ktsup{\mu}}{\kt}\right)+\frac{1}{3}\projkt^{\mu\nu}\kt \d u_\kt \right]\;.
\end{equation}
Substituting this expression into Eq.\ \eqref{eq:shear-eom} and contracting the result as species in Eq.\ \eqref{eq:dpi-components}, we obtain
\begin{subequations}\label{eq:shear-eom-k}
    \begin{align}
        & \left(\tau_{\pi}  \freq   - i\hbar \right)\frac{\d \pi_{\kt}}{\enp_\eq} 
     +  \frac{4\eta}{3\enp_\eq} \kt \d u_{\kt} =0\;,\label{eq:shear-eom-kt}\\
 & \left(\tau_{\pi}  \freq -i\hbar\right) \frac{\d \pi^\a_\perp}{\enp_\eq}  +
       \frac{\eta}{\enp_\eq}    {\kt} \delta u^\a_{\perp}=0\;,\label{eq:shear-eom-perp}
    \end{align}
\end{subequations}
and $\d\pi^{\mu\nu}_\perp = 0$.
Note that the transport coefficients $\eta$ and $\tau_\pi$ are evaluated in equilibrium.

As we did for Eq.\ \eqref{eq:em-cons-perp}, we contract Eq.\ \eqref{eq:shear-eom-perp} with $\epsilon^{\mu}{}_{\nu\rho\sigma}u^\rho_\eq\ktsup{\sigma}/\kt$ and utilize Eq.\ \eqref{eq:generic-cross}, to express it in terms of axial vectors.
This leads to
\begin{equation}
  \left(\tau_{\pi}  \freq -i\hbar\right) \frac{\d \pi^\mu_\chi}{\enp_\eq}  +
       \frac{\eta}{\enp_\eq} {\kt}    \delta u^\mu_{\chi}=0\;.\label{eq:shear-eom-cross}
\end{equation}

\subsection{Spin sector linear equations of motion}
Next, we consider the spin equations of motion, i.e., Eqs.\ \eqref{eq:ideal-spin-kappa} and \eqref{eq:ideal-spin-omega}.
After linearization, these equations take the forms
\begin{subequations}
    \begin{eqnarray}
     (A-B-C) \d\dot{\kappa}^{\langle\mu\rangle} &=& 
  E\epsilon^{\mu\nu\rho\sigma}u^\eq_\nu  \,\nabla_\rho \d\omega_\sigma + \hbar \Gamma^{(\kappa)}\left(\d\kappa^\mu + \d\varpi^{\mu\nu}u^\eq_\nu\right)-\inv{2}\hbar\Gamma^{(a)}\left(\nabla^\mu\d\beta + \b \d a^\mu\right)\;,\label{eq:ideal-spin-kappa-l}
  \\
  (A-2C)\d \dot{\omega}^{\langle\mu\rangle} &=&D\epsilon^{\mu\nu\rho\sigma}u^\eq_\nu  \,\nabla_\rho \d\omega_\sigma -\hbar \Gamma^{(\omega)}\left(\d\omega^\mu + \beta_\eq \d\kvort^\mu \right)\;,\label{eq:ideal-spin-omega-l}
\end{eqnarray}
\end{subequations}
where the coefficients $\{A,\cdots,E\}$, $\Gamma^{(\kappa)}$, $\Gamma^{(\omega)}$, and $\Gamma^{(a)}$ are evaluated in equilibrium.

We apply the replacements \eqref{eq:fourier-rep} to transform these equations into momentum space.
Then, we multiply Eq.\ \eqref{eq:ideal-spin-kappa-l} by $i/\Gamma^{(\kappa)}$ and Eq.\ \eqref{eq:ideal-spin-omega-l} by $-i/\Gamma^{(\omega)}$ to obtain
\begin{subequations}
    \begin{eqnarray}
    \left(\tau_\kappa \freq -i\hbar \right)\d{\kappa}^{\mu} -\mu_\kappa \epsilon^{\mu\nu\rho\sigma}u^\eq_\nu  \,\kt_\rho \d\omega_\sigma - i\hbar\d\varpi^{\mu\nu}u^\eq_\nu + \inv{2}\mu_a \left(\b_\eq \freq\d u^\mu + \ktsup{\mu} \d\beta\right) &=&0\;,\label{eq:ideal-spin-kappa-k}
  \\
   \left(\tau_\omega\freq - i\hbar\right)\d\omega^{\mu}+\mu_\omega\epsilon^{\mu\nu\rho\sigma}u^\eq_\nu  \,\kt_\rho \d\kappa_\sigma -i\hbar \beta_\eq \d\kvort^\mu &=&0\;,\label{eq:ideal-spin-omega-k}
\end{eqnarray}
\end{subequations}
where we have adopted the definitions from Ref.\ \cite{Wagner:2024fhf}:
\begin{equation}\label{eq:spin-relaxation-times}
   \tau_\kappa \equiv -\frac{A-B-C}{\hbar \Gamma^{(\kappa)}}\;,\qquad \mu_\kappa \equiv - \frac{E}{\hbar \Gamma^{(\kappa)}}\;,\qquad 
   \tau_\omega \equiv \frac{(A-2C)}{\hbar \Gamma^{(\omega)}}\;,\qquad \mu_\omega\equiv -\frac{D}{\hbar \Gamma^{(\omega)}}\;,
   \qquad \mu_a \equiv \frac{\Gamma^{(\kappa)}}{\Gamma^{(a)}}\;.
\end{equation}

To proceed with Eq.\ \eqref{eq:ideal-spin-kappa-k}, we need to compute the perturbation of the thermal vorticity $\d \varpi_{\mu\nu}$ in the momentum space.
By definition,
$$
    \d \varpi_{\mu\nu} = -\inv{2}\left(\partial_\mu \d \beta_\nu - \partial_\nu \d \beta_\mu\right)\;,
$$
which in momentum space transforms to
\begin{equation}\label{eq:varpi-k}
    \d \varpi_{\mu\nu} = -\frac{i}{2\hbar}\left(k_\mu \d \beta_\nu - k_\nu \d \beta_\mu\right)\;.
\end{equation}
Noting that
$$
\d\beta_\mu \equiv \d \left(\beta u_\mu\right) = \b_\eq\d u_\mu+u^\eq_\mu \d \b\;, 
$$
and substituting this relation in Eq.\ \eqref{eq:varpi-k}, we obtain
\begin{equation}\label{eq:tvort-k}
    \d \varpi_{\mu\nu} = -\frac{i}{2\hbar}\left[\d \beta\left(k_\mu  u^{\eq}_{\nu} - k_\nu  u^{\eq}_{\mu}\right)+\beta_\eq\left(k_\mu \d u_\nu - k_\nu \d u_\mu\right)\right]\;.
\end{equation}
Contracting this relation with $u^\eq_\nu$ and utilizing Eq.\ \eqref{eq:k-decomp} gives
\begin{equation}\label{eq:tvortu-k}
    \d\varpi^{\mu\nu}u^\eq_\nu = -\frac{i}{2\hbar}\left(\b_\eq\freq\d u^\mu - \d\b \ktsup{\mu}\right)\;.
\end{equation}
Substituting this result into Eq.\ \eqref{eq:ideal-spin-kappa-k} and contracting the result with $-\ktsub{\mu}/\kt$, we find
\begin{equation}\label{eq:ideal-spin-kappa-kt}
    \left(\tau_\kappa\freq - i\hbar\right) \d\kappa_\kt + \inv{2}\freq \b_\eq(1+\mu_a) \d u_\kt - \inv{2}(1-\mu_a)\kt \d\b =0\;.
\end{equation}

Similarly, contracting Eq.\ \eqref{eq:ideal-spin-kappa-k} with $\projkt^\a_\mu$ yields
\begin{equation}\label{eq:ideal-spin-kappa-perp}
    \left(\tau_\kappa\freq - i\hbar\right) \d\kappa^\a_\perp - \mu_\kappa \epsilon^{\a\b\sigma\rho}u^\eq_\b \kt_\sigma \d\omega^\perp_\rho + \inv{2}\freq \b_\eq(1+\mu_a) \d u^\a_\perp=0\;,
\end{equation}
where we have used Eq.\ \eqref{eq:generic-transverse-vector} with $V^\mu$ being $\d\kappa^\mu$ and $\d u^\mu$.

To align with our earlier choice of axial vector degrees of freedom, we contract Eq.\ \eqref{eq:ideal-spin-kappa-perp} with $\epsilon^{\mu}{}_{\a\b\lambda}u^\b_\eq\ktsup{\lambda}/\kt$ and employ the definition \eqref{eq:generic-cross} for $\d\kappa^\mu$ and $\d u^\mu$ to arrive at 
\begin{equation}\label{eq:ideal-spin-kappa-cross}
    \left(\tau_\kappa\freq - i\hbar\right) \d\kappa^\mu_\chi +\kt \mu_\kappa \d\omega^\mu_\perp + \inv{2}\freq \b_\eq(1+\mu_a) \d u^\mu_\chi=0\;.
\end{equation}

Equation \eqref{eq:ideal-spin-omega-k} includes the fluid vorticity vector $\d\kvort^\mu = -\tfrac{1}{2}\epsilon^{\mu\nu\a\b}u^\eq_\nu \nabla_\a \d u_\b$.
We apply the replacements \eqref{eq:fourier-rep} to transform it to momentum space and then use the definition \eqref{eq:generic-cross} for $\d u^\mu$, yielding
\begin{equation}\label{eq:kvort-k}
    \d\kvort^\mu = -\frac{i}{2\hbar}\kt \d u^\mu_\chi\;.
\end{equation}
Inserting this relation into Eq.\ \eqref{eq:ideal-spin-omega-k} gives
\begin{equation}\label{eq:ideal-spin-omega-k-2}
       \left(\tau_\omega\freq - i\hbar\right)\d\omega^{\mu}+\mu_\omega\kt \d\kappa^\mu_\chi -\inv{2} \beta_\eq \kt\d u^\mu_\chi =0\;.
\end{equation}
By contracting this equation with $-\ktsub{\mu}/\kt$, we realize that $\d\omega_\kt$ decouples from the rest of the equations,
\begin{equation}
       \left(\tau_\omega\freq - i\hbar\right)\d\omega_\kt =0\;.
\end{equation}
Finally, contracting Eq.\ \eqref{eq:ideal-spin-omega-k-2} with $\projkt^\mu_\a$ and relabeling the free index, we arrive at
\begin{equation}\label{eq:ideal-spin-omega-perp}
       \left(\tau_\omega\freq - i\hbar\right)\d\omega^{\mu}_\perp+\mu_\omega\kt \d\kappa^\mu_\chi -\inv{2} \beta_\eq \kt\d u^\mu_\chi =0\;.
\end{equation}

\subsection{Dispersion relations}
The equations derived in this section can be divided into two independent sets.
The first set comprises Eqs.\ \eqref{eq:em-cons-u}, \eqref{eq:em-cons-kt}, \eqref{eq:shear-eom-kt}, and \eqref{eq:ideal-spin-kappa-k}, in terms of the following set of dimensionless variables:
\begin{equation}\label{eq:sound-vars}
    \d X_{\rm sound} = \Big(\frac{\d\b}{\b_\eq},\,\d u_\kt,\, \frac{\d \pi_\kt}{\enp_\eq},\, \d\kappa_\kt\Big)^T\;.
\end{equation}
The subscript ``sound'' reflects that, excluding $ \d\kappa_\kt$, this set reduces to the sound channel of standard dissipative hydrodynamics.

The second set of equations consists of Eqs.\ \eqref{eq:em-cons-cross}, \eqref{eq:shear-eom-cross}, \eqref{eq:ideal-spin-kappa-cross}, and \eqref{eq:ideal-spin-omega-perp}, in terms of the following set of dimensionless variables:
\begin{equation}\label{eq:shear-vars}
    \d X_{\rm shear} = \Big(\d u_\chi,\, \frac{\d \pi_\chi}{\enp_\eq},\, \d\kappa^\mu_\chi, \d\omega^\mu_\perp \Big)^T\;.
\end{equation}
The subscript ``shear'' is used because, in the absence of $\d\kappa^\mu_\chi$ and $\d\omega^\mu_\perp$, this set corresponds to the shear modes of standard dissipative hydrodynamics.

\subsubsection{Sound Channel}

Representing Eqs.\ \eqref{eq:em-cons-u}, \eqref{eq:em-cons-kt}, \eqref{eq:shear-eom-kt}, and \eqref{eq:ideal-spin-kappa-k} in matrix form gives
\begin{eqnarray}
    M_{\rm sound} \d X_{\rm sound} &=& \left(\begin{array}{cccc}
\freq & \vs^2\kt & 0 & 0\\
\kt & \freq & - \kt & 0\\
0 & \tfrac{4}{3} \eta\kt/ \enp_\eq & \tau_{\pi}\freq -i \hbar & 0\\
- \tfrac{1}{2}(1-\mu_a) \kt & \tfrac{1}{2}(1+\mu_a) \freq & 0 &( \freq \tau_{\kappa}-i \hbar)/\beta_\eq
\end{array}\right)
\left(\begin{array}{c}
{\d\b}/{\b_\eq}\\
\d u_\kt\\ {\d \pi_\kt}/{\enp_\eq}\\ \d\kappa_\kt
\end{array}\right) = \vb{0}\;,
\end{eqnarray}
where $\vs^2 = \pdv*{P}{\varepsilon}$ is the squared speed of sound.
The determinant of $M_{\rm sound}$ decomposes into two distinct polynomials, each of them independently vanishing.
The first polynomial gives the characteristic equation
\begin{equation}
     (\freq^2  - \ktsup{2} v_s^2) \left(\tau_\pi\freq -i\hbar\right)  +\frac{4\eta}{3\enp} \ktsup{2}= 0\;,
\end{equation}
which is recognized as the standard sound channel in IS hydrodynamics without bulk viscosity.
The second polynomial corresponds to 
\begin{equation}
    \freq \tau_{\kappa}-i \hbar = 0\;,
\end{equation}
which describes the longitudinal spin modes and agrees with the static background results of Ref.\ \cite{Wagner:2024fhf}.

\subsubsection{Shear channel}

Equations \eqref{eq:em-cons-cross}, \eqref{eq:shear-eom-cross}, \eqref{eq:ideal-spin-kappa-cross}, and \eqref{eq:ideal-spin-omega-perp} take the following matrix form:
\begin{eqnarray}
    M_{\rm shear} \d X_{\rm shear} &=&    \left(\begin{array}{cccc}
\freq& - \kt & 0 & 0\\
\eta\kt/\enp_\eq &  \freq\tau_{\pi}{} -i \hbar & 0 & 0\\
\tfrac{1}{2} \beta (1+\mu_a) \freq& 0 &   \freq\tau_{\kappa}{} -i \hbar& \mu_{\kappa}{}\kt\\
- \tfrac{1}{2} \beta \freq& 0 & \mu_{\omega}{}\kt & -i \hbar + \freq\tau_{\omega}{}
\end{array}\right)
\left(\begin{array}{c}
     \d u_\chi\\ {\d \pi_\chi}/{\enp_\eq}
     \\ \d\kappa^\mu_\chi \\ \d\omega^\mu_\perp
\end{array}\right)=\vb{0}\;.
\end{eqnarray}
As with the sound channel, the determinant of the block diagonal matrix $M_{\rm shear}$ decomposes into a fluid and a spin part, rendering the spin waves independent of fluid perturbations.
The fluid part is the standard shear channel of IS hydrodynamics:
\begin{equation}
    \freq(\tau_\pi\freq-i\hbar) + \frac{\eta}{\enp_\eq}\ktsup{2} = 0\;.
\end{equation}
The spin part, once again, reproduces the results of Ref.\ \cite{Wagner:2024fhf}:
\begin{equation}\label{eq:spin-shear}
   \freq^2-i\hbar a\omega -\vsw^2\ktsup{2}-\hbar^2b = 0\;,
\end{equation}
where we have adopted the following definitions:
\begin{equation}
    a \equiv \frac{\tau_\kappa +\tau_\omega}{\tau_\kappa \tau_\omega} \;,\quad b\equiv \frac{1}{\tau_\kappa \tau_\omega}\;,\quad \vsw^2\equiv \frac{\mu_\kappa \mu_\omega}{\tau_\kappa \tau_\omega}\;.
\end{equation}

These results confirm that fluid perturbations do not influence the spin waves compared to a static background, consistent with the general proof in Sec.\ \ref{sec:linear-spin-hydro}.
They also extend the findings of Ref.\ \cite{Wagner:2024fhf} by demonstrating that the spin relaxation timescales remain the same even when the fluid is slightly perturbed.

\section{Ideal-spin hydrodynamics in a conformal Bjorken background}\label{sec:bjorken}

In this section, we study the evolution of the spin tensor within a conformal Bjorken background.
{
As shown in Appendix \ref{app:conformal}, conformal invariance imposes that
\begin{equation}
    T^{\mu}{}_\mu + D_\lambda \mcS_{\mu}{}^{\mu\lambda} = 0\;.
\end{equation}
However, this condition is only to be met in the classical limit, i.e., at zeroth order in $\hbar$, where it reduces to
\begin{equation}
    T^{\mu}{}_\mu = \order{\hbar^2}\;.
\end{equation}
Consequently, the fluid sector of the conformal background can be treated in the standard way, with the EOS 
\begin{equation}\label{eq:cf-eos}
  \varepsilon = 3P + \order{\hbar^2}\;.
\end{equation}
}

The Bjorken flow, a widely studied toy model for heavy-ion collisions, is often described using the so-called Milne coordinates $(\tau, x,y, \xi)$, which are defined in terms of Cartesian coordinates as follows:
\begin{equation}
    \tau=\sqrt{t^2-z^2}\;, \qquad \tanh\xi = \frac{z}{t}\;.
\end{equation}
The flat spacetime line element in these coordinates is given by
\begin{equation}
    \dd{s}^2 \equiv g_{\mu\nu} \dd{x}^\mu \dd{x}^\nu = \dd{\tau}^2 - \dd{x}^2 -\dd{y}^2 - \tau^2\dd{\xi}^2\;.
\end{equation}

In the Milne coordinates system, the Bjorken fluid four-velocity is $u^\mu = \left(1,\vb{0}\right)$, and all hydrodynamic quantities depend only on $\tau$.
Therefore, for any scalar quantity $X$, $\dot{X} = \dv*{X}{\tau}$ and $\nabla_\mu X = 0$.
The fluid vorticity and acceleration vanish, and the only nonzero components of the velocity four-gradient, $D_\mu u_\nu$, are the expansion scalar $\theta = 1/\tau$, and the shear tensor,
\begin{equation}\label{eq:bjorken-sigma}
    \sigma^\mu_\nu = \frac{2}{3\tau}{\rm diag}\left(0,-\inv{2}, -\inv{2}, 1\right)\;.
\end{equation}
Inserting this expression in Eq.\ \eqref{eq:shear-eom}, it follows that the shear-stress tensor $\pi^{\mu}_\nu$ has only one independent component, which we denote by $\pi(\tau)$.
Consequently, $\pi^{\mu}_\nu$ is expressed as
\begin{equation}\label{eq:bjorken-pi}
     \pi^\mu_\nu = \pi(\tau){\rm diag}\left(0,-\inv{2}, -\inv{2}, 1\right)\;.
\end{equation}
\subsection{Fluid sector: The hydrodynamic attractor}

In the Bjorken flow, the transverse part of the energy-momentum conservation \eqref{eq:gradb} is trivial.
We then substitute Eqs.\ \eqref{eq:bjorken-sigma} and \eqref{eq:bjorken-pi} into the fluid equations of motion \eqref{eq:bdot} and \eqref{eq:shear-eom}.
Using $\dot{\b}\pdv*{\varepsilon}{\beta}=\dot{\varepsilon}$, and {the conformal EOS \eqref{eq:cf-eos}}, these steps yield
\begin{subequations}
    \begin{eqnarray}
    \tau\dv{P}{\tau} &=& - \frac{4P}{3} + \frac{\pi(\tau)}{3}\;,\label{eq:bjorken-fluid-eom}
    \\
    \tau_\pi \dv{\pi}{\tau} + \pi(\tau)  &=& \frac{4\eta}{3} - \frac{4\tau_\pi}{3\tau}\left(1+C_{\pi\pi}\right)\pi(\tau)\;,\label{eq:bjorken-shear-eom}
\end{eqnarray}
\end{subequations}
where we have used $\d_{\pi\pi}=4/3\tau_\pi$ \cite{Denicol:2012cn}, and defined $C_{\pi\pi}\equiv 11\d_{\pi\pi}/(24\tau_\pi)$.

To combine these two equations into a single one, we use the conformal relation $P\sim T^4$, and then take the derivative of Eq.\ \eqref{eq:bjorken-fluid-eom} to determine $\pi'(\tau)$.
Substituting this result into Eq.\ \eqref{eq:bjorken-shear-eom} leads to the following second-order ordinary differential equation (ODE) in $T(\tau)$:
\begin{eqnarray}\label{eq:bjorken-master-T}
    \frac{\tau C_{\tau_{\pi}}}{T} \dv[2]{T}{\tau} + \frac{3\tau C_{\tau_{\pi}} }{T^2} \left(\dv{T}{\tau}\right)^2 + \left(\frac{11 C_{\tau_{\pi}} + 4 C_{\pi \pi} C_{\tau_{\pi}}}{3 T} + \tau\right) \dv{T}{\tau} + \frac{1}{3} T - \frac{4 C_{\eta}}{9 \tau} + \frac{4}{9 \tau} C_{\tau_{\pi}} \left(1 + C_{\pi \pi}\right) = 0\;.
\end{eqnarray}
Using a reparametrization introduced in Ref.\ \cite{Heller:2015dha}, we reduce the degree of this equation by 1. Instead of expressing this equation in terms of the variable $\tau$, we reformulate it using the dimensionless thermal time $w \equiv T\tau$.
Consequently, Eq.\ \eqref{eq:bjorken-master-T} transforms into a first-order nonlinear ODE, expressed in terms of the following function of $w$:
\begin{equation}\label{eq:hs-def}
    f = 1+\tau \frac{1}{T}\dv{T}{\tau} = \frac{\tau}{w} \dv{w}{\tau}\;.
\end{equation}
In IS-type theories of hydrodynamics, the function $f(w)$ is directly related to the pressure anisotropy:
\begin{equation}
    \mathcal{A} \equiv \frac{P_\perp-P_\parallel}{P_{\rm EQ}} =\frac{3\pi(\tau)}{2P} = 18\left(f(w)-\frac{2}{3}\right)\;,
\end{equation}
where the transverse and parallel pressures are defined via
\begin{equation}
    T^{\mu}_{\nu} = {\rm diag}\Big(3P,-P_\perp, -P_\perp, -P_\parallel\Big) + \order{\hbar^2}\;.
\end{equation}
Applying the reparametrization, Eq.\ \eqref{eq:bjorken-master-T} becomes
\begin{equation}\label{eq:bjorken-master-f}
     C_{\tau \pi}w f'(w) f(w)+4C_{\tau \pi}f(w)^2 + \left[w-\frac{16 C_{\tau \pi}}{3}\left(1-\frac{C_{\pi\pi}}{4}\right)\right]f(w) -\frac{4C_\eta}{3}
     +\frac{16C_{\tau_\pi}}{9}\left(1-\frac{C_{\pi\pi}}{2}\right)-\frac{2w}{3}= 0\;,
\end{equation}
where the following dimensionless coefficients are defined:
\begin{equation}
    C_\eta= \frac{\eta}{s}\;,\qquad C_{\tau \pi} = \tau_\pi T\;.
\end{equation}
Note that Eq.\ \eqref{eq:bjorken-master-f} reduces to Eq.\ (9) of Ref.\ \cite{Heller:2015dha} by setting $C_{\pi\pi}=0$.
To derive Eq.\ \eqref{eq:bjorken-master-f} from Eq.\ \eqref{eq:bjorken-master-T}, we employed Eq.\ \eqref{eq:hs-def} along with the following expression for the second-order derivative of $T(\tau)$:
\begin{equation}
    \dv[2]{T}{\tau} = \frac{w}{\tau^3}\left[2+f(w)^2+wf(w)f'(w)-3f(w)\right]\;.
\end{equation}

By setting the dissipative coefficients $C_{\tau \pi}$ and $C_\eta$ to zero, we find that for a perfect fluid $f=2/3$, which is a reflection of vanishing pressure anisotropy.
If only $C_{\tau \pi}$ is set to zero, the solution in the Navier-Stokes limit is obtained as
\begin{equation}\label{eq:f-ns}
    f(w) = \frac{2}{3} + \frac{4C_\eta}{9w}\;.    
\end{equation}

Equation \eqref{eq:bjorken-master-f} is numerically solvable, exhibiting the so-called attractor behavior: solutions initialized differently converge to a universal attractor solution at late thermal times.
Numerically, the attractor solution is found by setting $w=0$ in Eq.\ \eqref{eq:bjorken-master-f} and solving the resulting quadratic equation to find $f(0)$.
Among the two roots, the following is used as the initial condition for the numerical attractor:
\begin{equation}
      f(0) = \frac{2}{3} - \frac{C_{\pi\pi}}{6} + \frac{\sqrt{64C_\eta C_{\tau \pi}+(4C_{\tau \pi}C_{\pi \pi})^2}}{24C_{\tau \pi}}\;.
\end{equation}
The attractor can also be approximated using the so-called \textit{slow-roll approximation}.
For this purpose, we assume $f' = \epsilon_{\partial} f$ and expand Eq.\ \eqref{eq:bjorken-master-f} in powers of $\epsilon_{\partial}$.
Solving the equation obtained from the leading order in this expansion yields the following solution:
\begin{equation}\label{eq:f-slow-roll}
    f(w) = \frac{2}{3} - \frac{C_{\pi\pi}}{6} - \frac{w}{8C_{\tau_\pi}} + \frac{\sqrt{64C_\eta C_{\tau \pi}+(3w+4C_{\tau \pi}C_{\pi \pi})^2}}{24C_{\tau \pi}}\;.
\end{equation}
Expanding this solution in powers $1/w$ recovers the Navier-Stokes solution \eqref{eq:f-ns}.

\subsection{Spin sector}
We now turn to the ideal-spin dynamics equations, Eqs.\ \eqref{eq:ideal-spin-kappa} and \eqref{eq:ideal-spin-omega}.
In Milne coordinates, the electric and magnetic components of the spin potential can be parametrized as
\begin{equation}\label{eq:bjorken-spin-potential}
    \kappa^\mu =\left(0, \kappa_x(\tau),\kappa_y(\tau), \frac{\kappa_\eta(\tau)}{\tau}\right)\;,\qquad
    \omega^\mu =\left(0, \omega_x(\tau),\omega_y(\tau), \frac{\omega_\eta(\tau)}{\tau}\right)\;.
\end{equation}
The $x$ and $y$ components of these vectors break the rotational symmetry in the transverse ($xy$) plane. 
However, for illustrative purposes, we will disregard this symmetry breaking.

By inserting Eq.\ \eqref{eq:bjorken-spin-potential} into Eqs.\ \eqref{eq:ideal-spin-kappa} and \eqref{eq:ideal-spin-omega}, we arrive at the following equations for the spin components:
\begin{align}\label{eq:bjorken-spin-tau}
   & \tau_{\mathbf{x}} \left(\tau\dv{\tau} + 1\right)\mathbf{x}_\perp + \left(\d_{\mathbf{x}\beta}\frac{\tau}{T}\dv{T}{\tau} + \tau-\d_{\mathbf{x}D}\right) \mathbf{x}_\perp = 0\;,\\
   & \tau_{\mathbf{x}} \left(\tau\dv{\tau} + 1\right)\mathbf{x}_\eta +\left(\d_{\mathbf{x}\beta}\frac{\tau}{T}\dv{T}{\tau} + \tau\right)\mathbf{x}_\eta=0\;,
\end{align}
where $\mathbf{x}$ collectively denotes $\kappa$ and $\omega$, with $\mathbf{x}_\perp$ referring to their transverse ($x$ and $y$) components.
The coefficients appearing in this equation are defined as
\begin{equation}\label{eq:spin-coeffs}
    \d_{\kappa\beta} \equiv -\frac{T}{\hbar \Gamma^{(\kappa)}}\dv{T}(A-B-C)\;,\qquad
   \d_{\kappa \nabla} \equiv  -\frac{D}{\hbar \Gamma^{(\kappa)}}\;,\qquad
 \d_{\omega\beta} \equiv \frac{T}{\hbar \Gamma^{(\omega)}}\dv{T}(A-2C)\;,\qquad
   \d_{\omega \nabla} \equiv  \frac{E}{\hbar \Gamma^{(\omega)}}\;.
\end{equation}
Equation \eqref{eq:bjorken-spin-tau} reveals that, due to Bjorken symmetries, each component evolves independently of the others.

In heavy-ion collisions, the fluid is expected to start with unpolarized spin degrees of freedom, meaning that the spin potential is initially zero.
Due to the absence of source terms from the fluid’s thermal vorticity in Eq.\ \eqref{eq:bjorken-spin-tau}, an initially zero spin potential remains zero throughout the evolution within the Bjorken background.
Thus, we conclude the fluid remains unpolarized in the Bjorken flow. 

However, our purpose here is not to present a realistic description of heavy-ion collisions but to investigate how an initially finite spin potential relaxes to zero within the Bjorken background.

To this end, we scale the spin transport coefficients \eqref{eq:spin-coeffs} with the shear relaxation time, inspired by the approach in Ref.\ \cite{Wagner:2024fhf}, where the ratios of spin transport coefficients to the shear relaxation time were computed:
\begin{equation}\label{eq:new-coeffs}
    \tau_{\mathbf{x}} = C_{\tau_{\mathbf{x}}} \tau_{\pi}\;,\qquad \d_{{\mathbf{x}}\beta} = C_{{\mathbf{x}}\beta} \tau_{\pi}\;,\qquad
    \d_{{\mathbf{x}}\nabla} = C_{{\mathbf{x}}\nabla} \tau_{\pi}\;.
\end{equation}
Here, $C_{\tau_{\mathbf{x}}}$, $C_{{\mathbf{x}}\beta}$, and $C_{{\mathbf{x}}\nabla}$ are dimensionless parameters, which, for simplicity, we assume to be constants.
In other words, we assume that the spin transport coefficients depend solely on the powers of the temperature.

Substituting definitions \eqref{eq:new-coeffs} into Eq.\ \eqref{eq:bjorken-spin-tau} yields
\begin{align}\label{eq:bjorken-spin-tau-2}
   & C_{\tau_{\mathbf{x}}} \left(\tau\dv{\tau} + 1\right)\mathbf{x}_\perp + \left(C_{\mathbf{x}\beta}\frac{\tau}{T}\dv{T}{\tau} + \frac{\tau}{\tau\pi}-C_{\mathbf{x}\nabla}\right) \mathbf{x}_\perp = 0\;,\\
   & C_{\tau_{\mathbf{x}}} \left(\tau\dv{\tau} + 1\right)\mathbf{x}_\eta +\left(C_{\mathbf{x}\beta}\frac{\tau}{T}\dv{T}{\tau} + \frac{\tau}{\tau_\pi}\right)\mathbf{x}_\eta=0\;.
\end{align}
We then use the definition \eqref{eq:hs-def} to rewrite this equation in terms of the thermal time $w$ as
\begin{subequations}\label{eq:x-w-eom}
\begin{align}
   &  C_{\tau_{\mathbf{x}}}  \left(wf(w)\dv{w} + 1\right)\mathbf{x}_\perp - C_{{\mathbf{x}}\beta} \left[1-f(w)\right]\mathbf{x}_\perp + \left(\frac{w}{C_{\tau_\pi}}-C_{{\mathbf{x}}\nabla}\right)\mathbf{x}_\perp = 0\;,\\
   & C_{\tau_{\mathbf{x}}}  \left(wf(w)\dv{w} + 1\right)\mathbf{x}_\eta - C_{{\mathbf{x}}\beta} \left[1-f(w)\right]\mathbf{x}_\eta + \frac{w}{C_{\tau_\pi}}\mathbf{x}_\eta=0\;.\label{eq:x-w-eom-eta}
\end{align}
\end{subequations}
\begin{figure}
    \centering
    \includegraphics[width=0.5\linewidth]{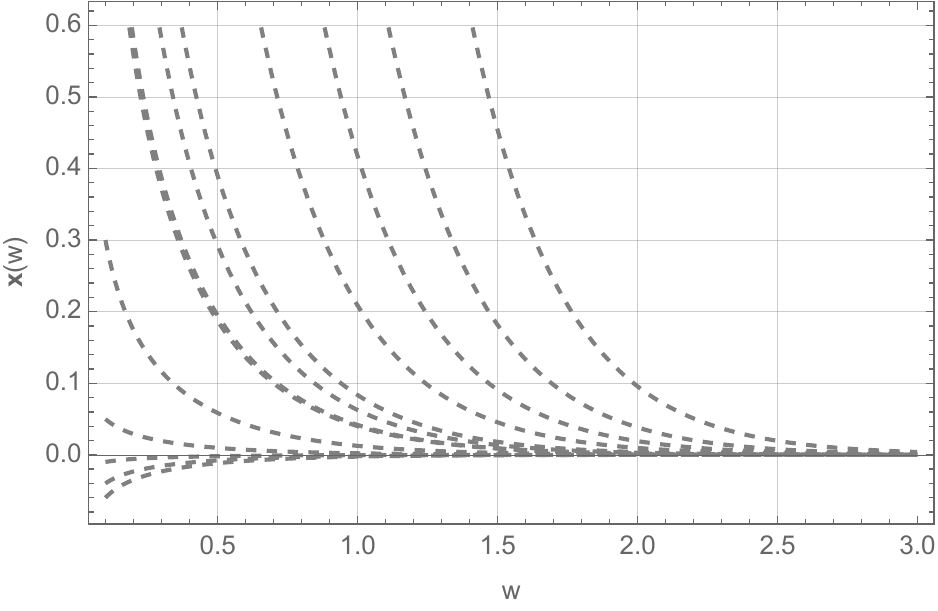}
    \caption{Relaxation of $\mathbf{x} \in \{\kappa_\eta,\omega_\eta\}$ to zero with Bjorken background.}
    \label{fig:bjorken-damp}
\end{figure}

After solving Eq.\ \eqref{eq:bjorken-master-f} for $f(w)$, we substitute it into Eq.\ \eqref{eq:x-w-eom} to determine $\kappa$ and $\omega$.
Figure \ref{fig:bjorken-damp} illustrates this procedure, where a nonzero $\mathbf{x}_\eta$ is assumed at $w=0$.
For $f(w)$, the slow-roll approximation \eqref{eq:f-slow-roll} is then substituted in Eq.\ \eqref{eq:x-w-eom-eta}.
To provide a concrete example, we arbitrarily set $C_{\tau_{\mathbf{x}}}=2$ and $C_{\mathbf{x}\beta}=1$, while the fluid transport coefficients $C_\eta = 1/(4\pi)$ and $C_{\tau_\pi} = (2-\log 2)/(2\pi)$ are taken from Ref.\ \cite{Bhattacharyya:2007vjd}.      
As anticipated, $\mathbf{x}_\eta$ rapidly relaxes to zero, demonstrating the relaxation dynamics of the spin potential in the Bjorken background.

To obtain the relaxation timescale, we examine Eq.\ \eqref{eq:x-w-eom} in late times, $w\gg 1$.
Substituting the slow-roll approximation \eqref{eq:f-slow-roll} for $f(w)$ into Eq.\ \eqref{eq:x-w-eom-eta}, we expand the equation in terms of $1/w$.
At the leading order, we obtain the following equation:
\begin{equation}
   \frac{2}{3}C_{\tau_{\mathbf{x}}}\left(w+\frac{2C_\eta}{3}\right)\dv{\mathbf{x}_\eta}{w} + \left[\frac{w}{C_{\tau_{\pi}}}+C_{\tau_\mathbf{x}}-\inv{3}C_{\mathbf{x}\beta}\right]\mathbf{x}_{\eta} =0\;.
\end{equation}
The solution to this equation at late times behaves as
\begin{equation}
    \mathbf{x}(w \gg 1) \propto \exp(-\frac{3\tau}{2\tau_\mathbf{x}})w^{\eta/(\enp \tau_\mathbf{x})}\;,
\end{equation}
where we have recovered the proper time $\tau$ and the coefficient $\tau_\mathbf{x}$.

This behavior aligns with the results of Ref.\ \cite{Wagner:2024fhf}, as in Sec.\ \ref{sec:example}, reaffirming that the spin relaxation time coefficients ($\tau_\kappa$ and $\tau_\omega$) govern the equilibration of spin potential to thermal vorticity, even within the Bjorken flow background.
In contrast to the linear regime, the electric and magnetic modes are uncoupled here, a consequence of Bjorken symmetries.
Notably, this resemblance between the relaxation dynamics in the linear regime and the Bjorken flow found here in spin dynamics mirrors observations in the fluid sector: deviations from the Navier-Stokes solution for $f(w)$, Eq.\ \eqref{eq:f-ns}, decay exponentially as $\exp(-{\tau}/{\tau_\pi})$ \cite{Heller:2015dha}.

\section{Conclusion and outlook}\label{sec:conclusion}

In this work, we investigated various aspects of semi-classical spin hydrodynamics, where hydrodynamic currents are derived from an expansion in $\hbar$, covering both flat and curved spacetimes. We first established definitions for angular-momentum currents to ensure their covariance under general coordinate transformations. This formulation was then extended to curved spacetimes, where we demonstrated that the conservation of the energy-momentum tensor requires specific modifications involving the Riemann curvature and spin tensors, in agreement with the findings of Ref.\ \cite{HEHL197655} but in torsionless metrics.
Additionally, we revised pseudo-gauge transformations to account for their applicability in curved spacetimes.

We introduced key assumptions that allowed us to study semi-classical spin hydrodynamics without explicitly invoking quantum kinetic theory. 
In the first order in $\hbar$, solving the semi-classical spin hydrodynamic equations involves first solving fluid equations of motion to determine the evolution of fluid fields. 
These solutions are then used as input to solve the spin equations of motion, ultimately yielding the spin potential on the so-called Cooper-Frye hypersurface.
On this hypersurface, the polarization of the final hadrons is calculated using an appropriate formula for the Pauli-Lubanski vector \cite{Weickgenannt:2022zxs}.

As a simple yet insightful example of this procedure, we studied the linearized semi-classical spin hydrodynamic equations in a general form. 
A key result of this analysis is that spin and fluid perturbations decouple in the linear regime, a direct consequence of truncating the equations at the first order in $\hbar$.
We analyzed the ideal-spin approximation in a dissipative fluid with shear viscosity as a concrete application, 
which confirmed our general conclusion---the decoupling of spin and fluid waves.
Moreover, this example generalizes prior findings \cite{Wagner:2024fhf}, showing that the damping of spin waves is governed exclusively by spin relaxation time coefficients, independent of linear fluid perturbations.
These insights provide a foundation for the linear treatment of spin hydrodynamic theories beyond the ideal-spin approximation \cite{Weickgenannt:2022zxs,Wagner:2024fhf}.

We also examined the application of the Gibbs stability criterion to semi-classical spin hydrodynamics, identifying its inherent limitations when equations are truncated at the first order in $\hbar$.
Consequently, the Gibbs stability criterion reduces to an approximation yielding stability conditions that apply exclusively to the fluid sector while neglecting spin effects, which cannot be consistently accounted for. This is because incorporating the spin terms without including the quantum corrections to the fluid sector leads to inconsistencies, such as erroneously predicting that the spin tensor and thermal vorticity vanish in equilibrium.
These corrections, which appear only in the second order in $\hbar$, remain beyond our current knowledge.
This observation, combined with our argument for slowly rotating reference equilibrium, signals the inherent anisotropy of the equilibrium state, which remains unaddressed in semi-classical spin hydrodynamics but has been previously explored in other approaches \cite{Ambrus:2019cvr,Becattini:2020qol,Palermo:2021hlf,Florkowski:2024bfw,Drogosz:2024gzv}.

Finally, we studied the spin dynamics on top of a conformal Bjorken flow background. 
{
Conformal symmetry, when correctly implemented via Weyl covariance, implies that the energy-momentum tensor satisfies \( T^{\mu}{}_\mu + D_\lambda \mcS_{\mu}{}^{\mu\lambda} = 0 \), reducing to tracelessness of the energy-momentum tensor only in the classical limit.
While Ref.\ \cite{Singh:2020rht} claimed that conformal invariance requires a traceless energy-momentum tensor and a totally antisymmetric spin tensor, we have shown that this conclusion is incorrect.
Our analysis, based on Weyl-covariant charge conservation, accounts for the curvature of conformally flat spacetimes and yields results that agree with the variational approach of Ref.\ \cite{Gallegos:2022jow}.}

Assuming conformal symmetry to hold solely in the classical limit, we applied it to the fluid sector while keeping the spin sector untouched, as it is second order in $\hbar$.
Utilizing the parametrization of Ref.\ \cite{Heller:2015dha}, we expressed equations of motion in terms of thermal time $w=T\tau$, enabling the use of attractor solutions for the fluid sector.
Using the slow-roll approximation for the attractor solution, we substituted these solutions into the spin equations of motion to study the late-time evolution of the spin tensor.
Our analysis showed that the damping of the spin potential, within a conformal Bjorken background, is governed by the spin relaxation time coefficients, paralleling the damping of spin waves in the linear regime.
This observation, like our results in the linear regime, extends the conclusions of Ref.\ \cite{Wagner:2024fhf} beyond the case of a fluid in hydrostatic equilibrium.

This work did not propose a new theory of spin hydrodynamics but rather examined the equilibration of the spin potential in more complicated settings than previously studied in Ref.\ \cite{Wagner:2024fhf}. 
It also aimed to set the stage for the development of a theory of general relativistic spin hydrodynamics.
The Bjorken flow background considered here possesses a zero thermal vorticity, and we demonstrated that perturbations of thermal vorticity around a homogeneous equilibrium do not influence the spin potential.
As a natural next step, we are studying the dynamics of the spin potential in a rigidly rotating fluid background.

The semi-classical spin hydrodynamic formulation presented in this work is applicable in curved spacetimes. 
However, quantum coupling with gravity emerges only at the second order in $\hbar$, potentially requiring invoking quantum kinetic theory in curved spacetime \cite{Fonarev:1993ht,Liu:2018xip,Liu:2020flb,Hayata:2020sqz}. 
Similarly, the coupling between spin and electromagnetic fields becomes relevant at the same order \cite{Weickgenannt:2019dks,Singh:2022ltu}. 
Addressing the inherent anisotropy of the equilibrium state remains a crucial step toward a more complete understanding of fluid-spin coupling.

\section*{Acknowledgments}
    The authors are supported by the Deutsche Forschungsgemeinschaft
	(DFG, German Research Foundation) through the Collaborative Research
	Center CRC-TR 211 ``Strong-interaction matter under extreme conditions''
	- project number 315477589 - TRR 211.
    M.\ S.\ is supported by the State of Hesse within the Research Cluster
	ELEMENTS (Project ID 500/10.006).
    A.\ C.\ and M.\ S.\ thank ECT* for support at the Workshop ``Spin and quantum features of QCD plasma" during which this work has been developed. 
    The authors thank D.\ H.\ Rischke for his support, guidance, and fruitful discussions.
J.\ S.\ and M.\ S.\ thank N.\ Weickgenannt for her help in the initial stages of this work.
The authors thank J.\ Lohr, Shi Pu, R.\ Ryblewski, R.\ Singh, Dong-Lin Wang, and D.\ Wagner for carefully reading the draft and providing valuable feedback.
M.\ S.\ also thanks A.\ Yarom for private discussions that initiated extending the formalism presented here to curved spacetimes.
A.\ C.\ and M.\ S.\ and thank the Galileo Galilei Institute for Theoretical Physics for the hospitality and the INFN for partial support during the completion of this work.

\section*{Data availability}
The data that support the findings of this article are openly available \cite{data}.
\appendix
\section{Slowly rotating equilibrium}\label{app:slow-rotation}

In this Appendix, we study the rigidly rotating equilibrium to provide a detailed explanation of the covariant definition of angular momentum and clarify the notion of a slowly rotating fluid.
As mentioned in Sec.\ \ref{sec:spin-hydro} and elaborated in Ref.\ \cite{Shokri:2023rpp}, a rigidly rotating equilibrium configuration is constructed by forming the thermal Killing vector $\beta^\star$ through the substitution of the generators of rotations into Eq.\ \eqref{eq:cov_beta_star} as
\begin{equation}\label{eq:bstar-rot}
    \beta^\star = b^\star  - \lambda^\star_r K^r \;,
\end{equation}
where $K^r$, with $r=1,2,3$, are given in Eq.\ \eqref{eq:gen-rot}.
The associated intensive parameters are defined as
\begin{eqnarray}\label{eq:rot-intesnive}
    \lambda_r^\star = -\frac{\Omega_r}{T_0}\;.
\end{eqnarray}
These parameters correspond to the partial derivative of entropy with respect to the total angular momentum component generated by the respective $K^r$ in global equilibrium,
\begin{equation}\label{eq:rot_to_entropy}
    \frac{\Omega_r}{T_0} = - \pdv{S}{J^r}\Big|_{\rm GTE}\;,
\end{equation}
where $J^r$ is defined in Eq.\ \eqref{eq:total-angular-momentum-integrated}.
This expression is the relativistic correspondent of the standard thermodynamic relation that conjugates angular velocity and angular momentum in a non-relativistic system (see, e.g.,  \cite{Landau_vol5} Ch.\ 26).

For instance, the generator of rotation around the $z$-axis is given by
\begin{equation}\label{eq:k-3}
    K^3_\mu  = \left(0, y, -x, 0\right)\;.
\end{equation}
The antisymmetric part of its gradient is expressed as
\begin{equation}
    D_{[\nu}K^3_{\mu]} = -\left(\d^1_\nu \d^2_\mu - \d^2_\nu \d^1_\mu\right)\;.
\end{equation}
By inserting $K^3_\mu$ and $D_{[\nu}K^3_{\mu]}$ into Eq.\ \eqref{eq:spin-orbit-cov}, the orbital and spin angular momentum along the $z$-axis are respectively obtained as
\begin{equation}
    -T^{\lambda\nu}K^3_\nu = x T^{\lambda y} - y T^{\lambda x} = L^{\lambda xy}\;, \qquad \frac{1}{2}\mcS^{\lambda \mu\nu} D_{[\nu}K^3_{\mu]} =  \mcS^{\lambda xy}\;.
\end{equation}
Summing these two terms, as prescribed by Eq.\ \eqref{eq:tot-angular-momentum}, yields the total angular momentum along the $z$-axis,
\begin{equation}
    J^{\mu xy} = -T^{\lambda\nu}K^3_\nu + \frac{1}{2}\mcS^{\lambda \mu\nu} D_{[\nu}K^3_{\mu]}\;.
\end{equation}
The thermal Killing vector and thermal vorticity are found to be
\begin{equation}
    \beta^{\star\mu} = \inv{T_0}\left(1,-y\, \Omega_3, x\, \Omega_3 , 0\right)\;,\qquad 
    \varpi^\star_{\mu\nu} = -\left(\d^1_\nu \d^2_\mu - \d^2_\nu \d^1_\mu\right)\;.
\end{equation}
As this example illustrates, instead of labeling $K$ with $r=1,2,3$, we can alternatively use $r=\{(yz),(zx),(xy)\}$, where here $r=3$ corresponds to $r=(xy)$.
This alternative labeling elucidates the connection between the Cartesian definition \eqref{eq:angular-momentum-rank3} and the covariant definition \eqref{eq:tot-angular-momentum} of angular momentum.

Let us now return to the general case of three generators of rotations in flat spacetime.
Substituting Eqs.\ \eqref{eq:bstar-rest} and \eqref{eq:rot-intesnive} into Eq.\ \eqref{eq:bstar-rot} yields the thermal Killing vector of rigidly rotating fluid,
\begin{eqnarray}\label{eq:bstar-rr}
    \beta^\star = \inv{T_0}\pdv{t} + \frac{\Omega_r}{T_0}K^r\;.
\end{eqnarray}
Writing this vector in the index notation, we compute $\partial_\nu\beta^\star_\mu$ using the definition \eqref{eq:gen-rot} in the form 
\begin{equation}\label{eq:gen-rot-index}
    K^r_\mu = \epsilon^{rij}g_{i\mu}x_j\;.  
\end{equation}
This gives rise to the thermal vorticity 
\begin{eqnarray}\label{eq:tvort-rr}
    \varpi^\star_{\mu\nu} = \frac{\Omega_{r}}{T_0} \epsilon^{rij}g_{i\mu}g_{j\nu}\;.
\end{eqnarray}
Plugging this expression into Eq.\ \eqref{eq:sr_beta_star} and rewriting the result in index-free notation, recovers Eq.\ \eqref{eq:bstar-rr}.

The Killing vectors $K^r$ are orthogonal to ${b}^\star$, i.e., ${b}^\star \cdot K^r = 0$, and satisfy
\begin{equation}
    K^m \cdot K^n = -\delta^{mn} \textbf{x}^2 + x^m x^n\;.
\end{equation}
Substituting Eq.\ \eqref{eq:bstar-rr} in the timeline condition $\beta_\mu^\star \beta^\mu{}^\star > 0$ and using these relations lead to
 \begin{equation}
 \label{eq:general-rotation-bound}
     \abs{\vb{\Omega}}^2 \abs{\vb{x}}^2 - \left(\vb{\Omega}\cdot\vb{x}\right)^2 < 1\;,
 \end{equation}
where $\vb{x}\equiv (x,y,z)$ and we have defined $\vb{\Omega}\equiv (\Omega_1,\Omega_2,\Omega_3)$.
This inequality generalizes the constraint on rigid rotation around the $z$-axis, $\Omega \rho < 1$,
with $\rho=\sqrt{x^2+y^2}$,\cite{Shokri:2023rpp}.
In the specific case of rigid rotation around the $z$-axis, the condition $\Omega \rho < 1$ ensures that the linear velocity remains subluminal, consistent with causality.
For a given $\Omega$, the \textit{causal size} of the system is defined as $R\equiv1/\Omega$, since at $\rho=R$ the linear velocity reaches the speed of light.
For the general case described by Eq.\ \eqref{eq:bstar-rr}, we similarly define a causal size $R\sim 1/\abs{\vb{\Omega}}$, using Eq.\ \eqref{eq:general-rotation-bound}.

The temperature is found from $T=1/\sqrt{\beta^\star\cdot\beta^\star}$ as $T = \gamma T_0$,
where $\gamma$ is the Lorentz factor,
 \begin{equation}\label{eq:gamma}
    \gamma = \inv{\sqrt{1-\vb{\Omega}^2 \vb{x}^2 + \left(\vb{\Omega}\cdot\vb{x}\right)^2}}\;.
\end{equation}
Subsequently, the four-velocity is obtained from $u = T\beta^\star$:
\begin{equation}\label{eq:four-velocity-rr}
    u = \gamma\left(\pdv{t} + {\Omega_r}K^r\right)\;.
\end{equation}

We now turn to the notion of a slowly rotating equilibrium.
Since the polarization is assumed to be small, the thermal vorticity of the reference equilibrium state must also be small.
In the LTE mapping, a large thermal vorticity of the reference equilibrium corresponds to a large spin potential of the fluid cell, allowing the fluid cells to develop significant polarization.
Thus, small polarization requires
\begin{equation}\label{eq:small-tvort}
    \abs{\vb{\Omega}} \ll T_0\;.
\end{equation}
However, this condition alone is insufficient to ensure small equilibrium gradients.

To understand the importance of small equilibrium gradients for hydrodynamic expansion, we consider the power counting commonly employed in standard approaches to hydrodynamics.
A crucial step is to identify the scalar, vector, and tensor gradients of hydrodynamic fields in combinations that vanish in global equilibrium.
For an uncharged fluid, the only vector term that is first order in gradients and vanishes in equilibrium is the following \cite{Kovtun:2012rj}:
\begin{equation}\label{eq:dissipative-vector}
    a_\mu - \frac{\nabla_\mu T}{T}\;.
\end{equation}
Now, let us examine the contribution of this term to the equation of motion \eqref{eq:fluid-eom-decomp},
\begin{equation}
    \left(\varepsilon+P\right)\left(a_\mu - \frac{\nabla_\mu T}{T}\right) = -\Pi a_\mu + \nabla_\mu \Pi + \cdots\;,
\end{equation}
where the shear stress tensor has been neglected for simplicity.
The standard argument is that the right-hand side of this equation is second-order in gradients and, thus, the left-hand side must also be second-order.
As a result, the vector \eqref{eq:dissipative-vector} does not appear in the Navier-Stokes theory for uncharged fluids.
However, if the equilibrium acceleration becomes large, this power-counting scheme breaks down, necessitating a revision of the first-order contributions.

Accordingly, we need to understand how to keep the temperature gradient, or, equivalently, the acceleration, small.
The equilibrium acceleration is found from $a_\mu \equiv  T \varpi_{\mu\nu}u^\nu$, which, using Eq.\ \eqref{eq:tvort-rr}, implies $a^\star_0 = 0$ and
\begin{equation}\label{eq:ai-rr}
    a^i = - \gamma\Omega_r \epsilon^{rij} u_j \;.
\end{equation}
This expression can be simplified further by noting that the spatial components of the four-velocity \eqref{eq:four-velocity-rr} are given by
\begin{equation}\label{eq:uj-rr}
    u^{j} =  \gamma\Omega_m \epsilon^{mjl}x_l\;,
\end{equation}
where Eq.\ \eqref{eq:gen-rot-index} has been used.
This relation implies that the components of the three-vector $\vb{v}$ are $v^{j}\equiv u^{j}/\gamma^\star=\Omega_m \epsilon^{mjl}x_l$ and, therefore, 
\begin{equation}
     \vb{\Omega}\cdot \vb{v} = 0\,.   
\end{equation}
By inserting Eq.\ \eqref{eq:uj-rr} into Eq.\ \eqref{eq:ai-rr}, we obtain
\begin{equation}
    a_{i} = -\gamma^2\left(\vb{\Omega}^2 x_i + \Omega_i \vb{x}\cdot\vb{\Omega}\right)\;.
\end{equation}
Using the equilibrium relation $\partial_\mu T = T a_\mu$, this yields
\begin{equation}\label{eq:gradi-t}
   \partial_i T = T_0\gamma^3\left(\vb{\Omega}^2 x_i + \Omega_i \vb{x}\cdot\vb{\Omega}\right)\;.
\end{equation}
This can be expressed in the Euclidean form as 
\begin{equation}\label{eq:grad-t}
   \va{\nabla} T = \gamma^2\left[-(\vb{\Omega}^2) \vb{x} + (\vb{x}\cdot\vb{\Omega})\vb{\Omega}\right]T\;.
\end{equation}

This expression reveals that even with a small $\abs{\vb{\Omega}}$, the temperature gradient can become large if the Lorentz factor $\gamma$ is large. According to Eq.\ \eqref{eq:gamma}, this can occur far from the center of rotation, where $\vb{x}=0$.
In the directions $x_i$, where the temperature gradient is nonzero, the system must have a finite size smaller than $R$ due to the constraint \eqref{eq:general-rotation-bound}.
We denote the typical scale of these sizes by $L_{\rm sys}$.
Close to the system's boundary, $\abs{\vb{x}} \sim L_{\rm sys}$, we substitute this into Eq.\ \eqref{eq:grad-t} to define the following dimensionless parameter:
\begin{equation}
    L_{\rm hydro}\abs{\vb{a}} = L_{\rm hydro}\abs{\frac{\va{\nabla} T}{T}} \sim \gamma^2 \left(\frac{ L_{\rm sys}}{R}\right)\left(\frac{L_{\rm hydro}}{R}\right)\;.
\end{equation}
To keep this quantity small, the following hierarchy must hold:
\begin{equation}
    L_{\rm hydro} \lesssim L_{\rm sys} \ll R\;. 
\end{equation}
The first part of this hierarchy, $L_{\rm hydro} \lesssim L_{\rm sys}$, reflects the trivial requirement that the size of fluid cells cannot exceed the size of the fluid itself.
The second part, $L_{\rm sys} \ll R$, implies that the typical fluid size must be significantly smaller than the maximum size allowed by causality.

In summary, we define a slowly rotating equilibrium with the following conditions:
\begin{enumerate}
    \item $\abs{\vb{\Omega}} \ll T_0$: Ensures that the thermal vorticity remains small, consistent with the small-polarization assumption.
    \item $L_{\rm sys}\abs{\vb{\Omega}}\ll 1$:  Prevents the emergence of large equilibrium gradients, preserving the validity of hydrodynamic expansion and maintaining approximate isotropy in equilibrium.
\end{enumerate}

\section{Pseudo-gauge transformation in curved spacetimes}\label{app:curved}
In this Appendix, we provide the details of the pseudo-gauge transformation \eqref{eq:cs-psgt} discussed in the main text.
For the reader's convenience, we restate the equations of motion \eqref{eq:cs-spin} here:
\begin{subequations}
   \begin{eqnarray}
    D_\mu T^{\mu\nu}&=&-\frac{1}{2}R^{\nu}{}_{\a\b\gamma}\mcS^{\a\b\gamma}\;,\label{eq:emt-cs}
    \\ 
    T^{[\a\b]}&=&-\frac{1}{2}D_\mu \mcS^{\mu\a\b}\;.\label{eq:cs-spin-dynamics}
\end{eqnarray}
\end{subequations}
Applying the standard pseudo-gauge transformation \eqref{eq:psgt} to Eq.\ \eqref{eq:cs-spin-dynamics} introduces an additional term, necessitating a revision of the transformation. 
Specifically, we find
\begin{equation}
    T^{[\mu\nu]}{}' = -\inv{2}D_\lambda \mcS^{\lambda\mu\nu}{}' -\inv{2}  D_\lambda D_\rho \Xi^{\mu\nu\lambda\rho}\;.
\end{equation}

This issue is resolved by noting that, like the superpotential $\Phi^{\lambda\mu\nu}$, the tensor $D_\rho \Xi^{\mu\nu\lambda\rho}$ appearing in the transformation of the spin tensor in Eq.\ \eqref{eq:psgt} is also antisymmetric in the first two indices.
Consequently, it can be absorbed into the definition of $\Phi^{\lambda\mu\nu}$.
We then define
\begin{equation}\label{eq:corrected-Z-1}
    \tilde{Z}^{\lambda\mu\nu} = \frac{1}{2}\left(\tilde{\Phi}^{\lambda\mu\nu} - \tilde{\Phi}^{\mu\lambda\nu} - \tilde{\Phi}^{\nu\lambda\mu}\right)\;,
\end{equation}
where
\begin{equation}\label{eq:corrected-super-pot-1}
   \tilde{\Phi}^{\lambda\mu\nu} \equiv \Phi^{\lambda\mu\nu} + D_\rho \Xi^{\mu\nu\lambda\rho}\;.
\end{equation}
This adjustment redefines the pseudo-gauge transformation \eqref{eq:psgt} as follows:
\begin{equation}\label{eq:cs-psgt-1}
    T^{\mu\nu}{}' = T^{\mu\nu} + D_\lambda \tilde{Z}^{\lambda\mu\nu}\;,\qquad S^{\lambda\mu\nu}{}' = S^{\lambda\mu\nu} - \tilde{\Phi}^{\lambda\mu\nu}\;.
\end{equation}
Using Eq.\ \eqref{eq:corrected-Z-1}, we observe that
\begin{equation}
  \tilde{\Phi}^{\lambda\mu\nu} = 2\Tilde{Z}^{\lambda[\mu\nu]}\;.
\end{equation}
Applying the updated transformation \eqref{eq:cs-psgt-1} to Eq.\ \eqref{eq:cs-spin-dynamics} and employing this relation, we find 
\begin{equation}
    T^{[\mu\nu]}{}' = -\inv{2} D_\lambda S^{\lambda\mu\nu}{}'\;,
\end{equation}
implying that Eq.\ \eqref{eq:cs-spin-dynamics} is covariant under the revised transformation\eqref{eq:cs-psgt-1}.

We now turn to Eq.\ \eqref{eq:emt-cs}, which, under transformation \eqref{eq:cs-psgt-1}, transforms as
\begin{equation}\label{eq:d-tmn-prime-1}
     D_\mu T^{\mu\nu}{}' = D_\mu T^{\mu\nu} + D_\mu D_\lambda \tilde{Z}^{\lambda\mu\nu} \;.
\end{equation}
We now insert $\tilde{Z}^{\lambda\mu\nu}$ into the Ricci identity and use the fact that $\tilde{Z}^{\lambda\mu\nu}$ is antisymmetric in its first two indices.
This yields
\begin{equation}
     D_\mu D_\lambda \tilde{Z}^{\lambda\mu\nu} =  -\inv{2}R^{\nu}{}_{\lambda\a\b}\tilde{Z}^{\a\b\lambda}\;.
\end{equation}
Substituting this relation in Eq.\ \eqref{eq:d-tmn-prime-1}, we arrive at 
\begin{align}\label{eq:d-tmn-prime-2}
    D_\mu T^{\mu\nu}{}' = D_\mu T^{\mu\nu} -\inv{2}R^{\nu}{}_{\lambda\a\b}\tilde{Z}^{\a\b\lambda}\;.
\end{align}
Next, we observe that the cyclic property of $R^{\nu}{}_{\lambda\a\b}$, the first Bianchi identity, can be alternatively applied to $\tilde{Z}^{\a\b\lambda}$.
Using this alongside the definition \eqref{eq:corrected-Z-1}, we find
\begin{equation*}
    R^{\nu}{}_{\lambda\a\b}\tilde{Z}^{\a\b\lambda} = - R^{\nu}{}_{\lambda\a\b}\left(\tilde{Z}^{\a\b\lambda} + \tilde{Z}^{\b\lambda\a}\right)
    =  -R^{\nu}{}_{\lambda\a\b}\tilde{\Phi}^{\lambda\a\b}\;.
\end{equation*}
Plugging this relation into Eq.\ \eqref{eq:d-tmn-prime-2}, we find
\begin{align*}
    D_\mu T^{\mu\nu}{}' &=  D_\mu T^{\mu\nu} +\inv{2}R^{\nu}{}_{\lambda\a\b}\tilde{\Phi}^{\lambda\a\beta}{}'
    \\ &= \inv{2}R^{\nu}{}_{\lambda\a\b}\left(-\mcS^{\lambda\a\beta}+\tilde{\Phi}^{\lambda\a\beta}-D_\rho\Xi^{\mu\nu\lambda\rho}\right)
    \\ &= -\inv{2}R^{\nu}{}_{\lambda\a\b}\mcS^{\lambda\a\beta}{}'\;,
\end{align*}
where we have used the transformation of the spin tensor under the pseudo-gauge transformation \eqref{eq:cs-psgt-1}.
Thus, we conclude that Eq.\ \eqref{eq:emt-cs} also transforms covariantly under the pseudo-gauge transformations \eqref{eq:cs-psgt-1}.

The redefinition of the superpotential $\Phi^{\lambda\mu\nu}$ in Eq.\ \eqref{eq:corrected-super-pot-1} is not restricted to adding the divergence of a rank-4 tensor.
For example, one can also include a rank-6 tensor as
\begin{gather}
     \tilde{\Phi}^{\lambda\mu\nu}= \Phi^{\lambda\mu\nu} + D_\rho \Xi^{\mu\nu\lambda\rho} + D_\tau D_\sigma D_\rho \Psi^{\mu\nu\lambda\rho\sigma\tau}\;,
\end{gather}
where $\Psi^{(\mu\nu)\lambda\rho\sigma\tau} = \Psi^{\mu\nu(\lambda\rho)\sigma\tau} = \Psi^{\mu\nu\lambda\rho(\sigma\tau)} = 0$.
Adding arbitrary terms in this form does not alter the conclusions presented earlier, as these arguments rely solely on the antisymmetry of $\tilde{\Phi}^{\lambda\mu\nu}$ in the last two indices, the first Bianchi identity, and the definition \eqref{eq:corrected-Z-1}.
Thus, we conclude that under the pseudo-gauge transformations \eqref{eq:cs-psgt} with the corrected superpotential \eqref{eq:cs-superpotential} of the main text, the equations of motion \eqref{eq:cs-spin} remain covariant.

\section{Conformal spin hydrodynamics}\label{app:conformal}
In this Appendix, we study the equations of spin hydrodynamics under conformal transformations.
A conformal transformation corresponds to a local rescaling of the metric tensor \cite{Carroll:1989vb, Wald:1984rg}
\begin{equation}\label{eq:conformal}
    g_{\mu\nu}(x) \longrightarrow \Tilde{g}_{\mu\nu}(x) = \Omega^2(x) g_{\mu\nu}(x)\;,
\end{equation}
where $\Omega(x)$ is a smooth, strictly positive function known as the conformal factor.
The determinant and the inverse metric transform as $\sqrt{-\Tilde{g}} = \Omega^4(x)\sqrt{-g}$ and $\Tilde{g}^{\mu\nu} = \Omega(x)^{-2}{g}^{\mu\nu}$, respectively.

We denote the covariant derivative associated with the original metric $g_{\mu\nu}$ by $D$ and the one corresponding to the \textit{conformally related metric} $\Tilde{g}_{\mu\nu}$ by $\Tilde{D}$.
To define $\Tilde{D}$ , we require the Christoffel symbols of $\Tilde{g}_{\mu\nu}$, which are related to those of $g_{\mu\nu}$ by
\begin{equation}
    \Tilde{\Gamma}^\rho_{\mu\nu} = {\Gamma}^\rho_{\mu\nu} + C^\rho_{\mu\nu} \;,
\end{equation}
where $C^\rho_{\mu\nu}$ is a tensor given by
\begin{align}
    C^\rho_{\mu\nu} =  \frac{1}{\Omega} \left( \delta^\rho_\nu D_\mu \Omega + \delta^\rho_\mu D_\nu \Omega - g_{\mu\nu} g^{\rho\sigma} D_\sigma \Omega \right)\;.
\end{align}
A tensor $\Psi$ of arbitrary rank is said to be a Weyl tensor with conformal weight $s$, where $s$ is a real number, if it transforms as
\begin{equation}\label{eq:weyl-tensor}
    \tilde{\Psi} = \Omega^s \Psi\;.
\end{equation}
An equation is \textit{Weyl-covariant} if, given a solution $\Psi$ in the original metric, $\Tilde{\Psi}$ is also a solution of the transformed equation in the conformally related metric.

If the action is invariant under the conformal transformation \eqref{eq:conformal}, the Hilbert energy-momentum tensor \eqref{eq:hilbert} is Weyl tensor transforming as $\tilde{T}^{\mu\nu}_{\rm H} = \Omega^{-6}{T}^{\mu\nu}_{\rm H}$.
Assuming that the Hilbert and Belifante tensors coincide, this suggests that the energy-momentum tensor defined in an arbitrary pseudo-gauge, related to the Belifante tensor via Eq.\ \eqref{eq:belifante}, cannot be a Weyl tensor. 
To find the correct transformations of the energy-momentum and spin tensors, we postulate that $J^\mu$ in Eq.\ \eqref{eq:cons_current-div} is a Weyl vector transforming as \footnote{Here, we have dropped the index $I$ for notation's simplicity.}
\begin{equation}\label{eq:cf-charge-current}
      \Tilde{J}^{\mu}  = \Omega^{-4}{J}^{\mu}\;.
\end{equation}
Consequently, Eq.\ \eqref{eq:cons_current-div} is Weyl-covariant,
\begin{equation}\label{eq:j-weight}
    \tilde{D}_\mu \tilde{J}^{\mu} = 0\;.
\end{equation}

Under a conformal transformation, a Killing vector $K$ of the metric $g$ becomes a conformal Killing vector $\tilde{K}$ of the metric $\tilde{g}$ satisfying
\begin{equation}\label{eq:cf-killing-w}
        \tilde{K}_\mu = \Omega^{2} K_\mu\;,
 \end{equation}
 which satisfy \cite{Zee:2013dea}
 \begin{equation}\label{eq:cf-killing}
     \tilde{D}_\mu\tilde{K}_\nu + \tilde{D}_\nu\tilde{K}_\mu = \inv{2} \tilde{g}_{\mu\nu} \tilde{D}\cdot \tilde{K}\;.
 \end{equation}
 Equation \eqref{eq:cf-killing} shows that, unlike a Killing vector, a conformal Killing vector has a nonzero divergence given by
 \begin{equation}\label{eq:cf-div}
      \tilde{D}\cdot \tilde{K} = 4  K \cdot \partial\ln \Omega
 \end{equation}
 As a consequence,  
    \begin{equation}\label{eq:cf-asym-grad}
        \tilde{D}_{[\mu}\tilde{K}_{\nu]} = 2\Omega K_{[\nu}\partial_{\mu]}\Omega - \Omega^2 D_{[\nu} K_{\mu]}\;.
    \end{equation}

 Using this equation and the cyclic property of the Riemann tensor, we obtain the following identity for the second-order gradient of a conformal Killing vector:
 \begin{equation}\label{eq:cf-killing-id}
      \tilde{K}_{\a|\b\gamma} =  \tilde{R}^{\nu}{}_{\gamma\b\a}\tilde{K}_\nu +\tilde{g}_{\a[\b}\tilde{D}_{\gamma]} \tilde{D}\cdot  \tilde{K}\;,
 \end{equation}
where $\tilde{K}_{\a|\mu} \equiv \tilde{D}_\mu \tilde{K}_\a$.

Now to determine $\tilde{J}$ in terms of $\tilde{T^{\mu\nu}}$ and $\tilde{K}$, we first write
\begin{align}\label{eq:j-t-step-1}
     -  \tilde{D}_\mu\left(\tilde{T}^{\mu\nu}\tilde{K}_\nu\right) &= -\tilde{K}_\nu \tilde{D}_\mu \tilde{T}^{\mu\nu} - \tilde{T}^{(\mu\nu)}\tilde{D}_{(\mu}\tilde{K}_{\nu)}- \tilde{T}^{[\mu\nu]}\tilde{D}_{[\mu}\tilde{K}_{\nu]}\;.
\end{align}
\textit{Defining} the tensor $\tilde{\mcS}^{\mu\a\b}$ via 
\begin{equation}\label{eq:cf-spin-tensor}
    \tilde{D}_\mu\tilde{\mcS}^{\mu\a\b} = 2\Tilde{T}^{[\b\a]}\;,
\end{equation}
Eq.\ \eqref{eq:j-t-step-1} is expressed as
\begin{align}\label{eq:j-t-step-2}
     -  \tilde{D}_\mu\left(\tilde{T}^{\mu\nu}\tilde{K}_\nu\right) 
     &= -\tilde{K}_\nu \tilde{D}_\mu \tilde{T}^{\mu\nu} - \inv{4}\tilde{T}^{\mu}{}_\mu\tilde{D}\cdot\tilde{K}+\inv{2}\tilde{D}_\mu\tilde{\mcS}^{\mu\a\b}\tilde{D}_\a\tilde{K}_{\b}\;,
\end{align}
Using the Leibniz rule, Eqs.\ \eqref{eq:cf-div} and \eqref{eq:cf-killing-id}, the last term in Eq.\ \eqref{eq:j-t-step-2} can be rewritten as
\begin{align}
    \inv{2}\tilde{D}_\mu\tilde{\mcS}^{\mu\a\b}\tilde{D}_\a\tilde{K}_{\b} &= \inv{2}\tilde{D}_\mu\left(\tilde{\mcS}^{\mu\a\b}\tilde{D}_\a\tilde{K}_{\b}\right) - \inv{2}\tilde{\mcS}^{\gamma\b\a}\tilde{R}^{\nu}{}_{\gamma\b\a}\tilde{K}_\nu -\inv{2}\tilde{D}_{\mu}\left(\inv{2}\tilde{\mcS}_{\a}{}^{\a\mu} \tilde{D}\cdot  \tilde{K}\right)+\inv{4}\tilde{D}_{\mu}\tilde{\mcS}_{\a}{}^{\a\mu} \tilde{D}\cdot  \tilde{K}\;.
\end{align}
Inserting this relation back into Eq.\ \eqref{eq:j-t-step-1}, we obtain
\begin{align}\label{eq:cf-ansatz-motif}
  -  \tilde{D}_\mu\left(\tilde{T}^{\mu\nu}\tilde{K}_\nu-\inv{2}\tilde{\mcS}^{\mu\a\b} \tilde{D}_\b \tilde{K}_\a- \inv{4}\tilde{\mcS}_\a{}^{\a\mu}\tilde{D}\cdot\tilde{K}\right) &= -\tilde{K}_\nu \left(\tilde{D}_\mu \tilde{T}^{\mu\nu} + \inv{2}\tilde{\mcS}^{\lambda\a\b}\tilde{R}^{\nu}{}_{\lambda\a\b}\right)\n&\hspace{-3cm}
  - \inv{4}\left(\tilde{T}^\a{}_\a + \tilde{D}_\a\tilde{\mcS}_{\b}{}^{\b\a} \right)\tilde{D}\cdot \tilde{K}\;.
\end{align}
For the Belifante energy-momentum tensor $\tilde{T}^{\mu\nu}_{\rm B}$, the right-hand side of this equation vanishes assuming $T^{\mu\nu}_{\rm B} = T^{\mu\nu}_{\rm H}$ is a Weyl tensor with vanishing trace.
The Belifante energy-momentum tensor is related to the energy-momentum tensor in an arbitrary pseudo-gauge through Eq.\ \eqref{eq:belifante}. 
Contracting both sides of Eq.\ \eqref{eq:belifante} with $g_{\mu\nu}$, and setting the left-hand side to zero, we obtain
\begin{equation}
    T^{\mu}{}_\mu + D_\lambda \mcS_{\mu}{}^{\mu\lambda} = 0\;.
\end{equation}
Inspired by this, we postulate that this equation and the equation of motion \eqref{eq:cs-emt} are Weyl-covariant, i.e., 
\begin{subequations}\label{eq:cf-eom}
    \begin{align}
        \tilde{D}_\mu \tilde{T}^{\mu\nu} + \inv{2}\tilde{S}^{\lambda\a\b}\tilde{R}^{\nu}{}_{\lambda\a\b} = 0\;, \\
        \tilde{T}^\a{}_\a + \tilde{D}_\a\tilde{S}_{\b}{}^{\b\a}  = 0\;.
    \end{align}
\end{subequations}
Using these equations in Eq.\ \eqref{eq:cf-ansatz-motif}, we define
 \begin{equation}\label{eq:j-ansatz}
         \tilde{J}^{\mu} = - \Tilde{T}^{\mu\nu}\Tilde{K}_{\nu} + \inv{2} \tilde{\mcS}^{\mu\a\b}\tilde{D}_{[\b}\tilde{K}_{\a]}-\inv{4}\tilde{\mcS}_{\a}{}{}^{\a\mu}\tilde{D}\cdot  \tilde{K}\;,
    \end{equation}
which satisfies $ \tilde{D}_\mu \tilde{J}^{\mu} = 0$.

Now, we derive the transformations of the energy-momentum and spin tensors by comparing Eq.\ \eqref{eq:j-ansatz} and Eq.\ \eqref{eq:cf-charge-current}.
We start by writing Eq.\ \eqref{eq:j-ansatz} as
    \begin{align}
       \Tilde{J}^{\mu}  &= \left[- \Omega^2 \Tilde{T}^{\mu\nu}-\inv{\Omega}\tilde{\mcS}_{\a}{}^{\a\mu}\partial^\nu\Omega+ \Omega \tilde{\mcS}^{\mu\nu\a}\partial_{\a}\Omega\right]{K}_{\nu} - \inv{2} \tilde{\mcS}^{\mu\a\b} \Omega^2 D_\a K_\b\;,
    \end{align}
where we have used Eqs.\ \eqref{eq:cf-killing-w}, \eqref{eq:cf-div} and \eqref{eq:cf-asym-grad}.
Equating this relation to the postulated transformation \eqref{eq:cf-charge-current}, we find 
\begin{align}
    \left(\Omega^{-4}{T}^{\mu\nu}- \Omega^2 \Tilde{T}^{\mu\nu}-\inv{\Omega}\tilde{\mcS}_{\a}{}^{\a\mu}\partial^\nu\Omega+ \Omega \tilde{\mcS}^{\mu\nu\a}\partial_{\a}\Omega\right){K}_{\nu} - \inv{2}\left(-\Omega^{-4} {\mcS}^{\mu\a\b}  + \tilde{\mcS}^{\mu\a\b} \Omega^2\right) D_\a K_\b = 0\;.
\end{align}
The term coupled to $D_\a K_\b $ must vanish independently, giving rise to
\begin{equation}
   \tilde{\mcS}^{\mu\a\b} = \Omega^{-6}  {\mcS}^{\mu\a\b}\;.
\end{equation}
The energy-momentum tensor, on the other hand, must transform as 
\begin{align}\label{eq:emt-cf-trans}
     \Tilde{T}^{\mu\nu} &=\Omega^{-6}\left({T}^{\mu\nu}-{\mcS}_{\a}{}^{\a\mu}\partial^\nu\ln\Omega+  {\mcS}^{\mu\nu\a}\partial_{\a}\ln\Omega\right)\;.
\end{align}
This is consistent with the Belifante energy-momentum tensor being a Weyl tensor and the results of Ref.\ \cite{Gallegos:2022jow}
found for the canonical tensors using a variational approach.
We also note that in a pseudo-gauge where ${\mcS}^{\mu\a\b}$ is totally antisymmetric, the energy-momentum tensor is traceless, albeit not a Weyl tensor.

Equations \eqref{eq:cf-eom} and \eqref{eq:cf-spin-tensor} are Weyl covariant by construction, and a direct calculation admits this.
As an example, we briefly comment on Eq.\ \eqref{eq:cf-spin-tensor}.
The divergence of $\tilde{\mcS}^{\mu\a\b}$ transforms as follows:
\begin{equation}
    \tilde{D}_\mu \tilde{\mcS}^{\mu\a\b} = \Omega^{-6}\left( {D}_\mu {\mcS}^{\mu\a\b} + 2 \mcS_{\mu}{}^{\mu[\a}\partial^{\b]}\ln\Omega -  2 \mcS^{[\a\b]\mu}\partial_{\mu}\ln\Omega\right)\;.
\end{equation}
This is consistent with the Weyl covariance of Eq.\ \eqref{eq:spin-dynamics}, noting that the antisymmetric part of the energy-momentum tensor  transforms as 
\begin{align}\label{eq:emt-cf-trans}
     \Tilde{T}^{[\a\b]} &=\Omega^{-6}\left({T}^{[\a\b]}-{\mcS}_{\mu}{}^{\mu[\a}\partial^{\b]}\ln\Omega+  {\mcS}^{[\a\b]\mu}\partial_{\mu}\ln\Omega\right)\;.
\end{align}

\bibliography{main}
\end{document}